\newif\ifhasbib

\hasbibtrue

\documentclass[twocolumn]{aastex61}

\submitjournal{ApJ}
\accepted{25 Aug 2017}

\shorttitle{Mdot-Mdisk-Mstar}
\shortauthors{Mulders et al.}

\usepackage{xspace} 

\newcommand{\Alma}{\texttt{ALMA}\xspace}
\newcommand{\Xshooter}{\texttt{X-Shooter}\xspace}
\newcommand{\linmix}{\texttt{linmix}\xspace}

\newcommand{\Tdust}{\ensuremath{T_{\rm dust}}\xspace}
\newcommand{\Mdust}{\ensuremath{M_{\rm dust}}\xspace}
\newcommand{\Mdisk}{\ensuremath{M_{\rm disk}}\xspace}
\newcommand{\Macc}{\ensuremath{\dot{M}_{\rm acc}}\xspace}
\newcommand{\Mstar}{\ensuremath{M_\star}\xspace}

\newcommand{\logMdust}{\ensuremath{\log \left(\frac{M_{\rm dust}}{M_\oplus}\right)}\xspace}
\newcommand{\logMacc}{\ensuremath{\log \left(\frac{\dot{M}_{\rm acc}}{M_\odot/{\rm yr}}\right)}\xspace}
\newcommand{\logMstar}{\ensuremath{\log\left(\frac{M_\star}{M_\odot}\right)}\xspace}

\newcommand{\Teff}{\ensuremath{T_{\rm eff}}\xspace}

\newcommand{\Msun}{\ensuremath{M_\odot}\xspace}

\newcommand{\DD}{\ensuremath{\Delta M_{\rm dust}}\xspace}
\newcommand{\DA}{\ensuremath{\Delta \dot{M}_{\rm acc}}\xspace}

\newcommand{\fgtd}{\ensuremath{f_{\rm gtd}}\xspace}
\newcommand{\facc}{\ensuremath{f_{\rm acc}}\xspace}

\newcommand{\visa}{{\rm\texttt{vis1}}\xspace}
\newcommand{\visb}{{\rm\texttt{vis2}}\xspace}
\newcommand{\visc}{{\rm\texttt{vis3}}\xspace}

\begin{document}

\title{Constraints from Dust Mass and Mass Accretion Rate Measurements on Angular Momentum Transport in Protoplanetary Disks}

\correspondingauthor{Gijs D. Mulders}
\email{mulders@lpl.arizona.edu}

\author{Gijs D. Mulders}
\affil{Lunar and Planetary Laboratory, The University of Arizona, Tucson, AZ 85721, USA}
\affil{Earths in Other Solar Systems Team, NASA Nexus for Exoplanet System Science}

\author{Ilaria Pascucci}
\affiliation{Lunar and Planetary Laboratory, The University of Arizona, Tucson, AZ 85721, USA}
\affiliation{Earths in Other Solar Systems Team, NASA Nexus for Exoplanet System Science}

\author{Carlo F. Manara}
\affiliation{Scientific Support Office, Directorate of Science, European Space Research and Technology Centre (ESA/ESTEC), Keplerlaan 1, 2201 AZ Noordwijk, The Netherlands}

\author{Leonardo Testi}
\affiliation{European Southern Observatory, Karl-Schwarzschild-Strasse 2, D-85748 Garching bei M{\"u}nchen, Germany}
\affiliation{INAF-Arcetri, Largo E. Fermi 5, I-50125 Firenze, Italy}
\affiliation{Gothenburg Center for Advance Studies in Science and Technology, Chalmers University of Technology and University of Gothenburg, SE-412 96 Gothenburg, Sweden}

\author{Gregory J. Herczeg}
\affiliation{Kavli Institute for Astronomy and Astrophysics, Peking University, Yi He Yuan Lu 5, Haidian Qu, 100871 Beijing, China}

\author{Thomas Henning}
\affiliation{Max Planck Institute for Astronomy, K{\"o}nigstuhl 17, D-69117 Heidelberg, Germany}

\author{Subhanjoy Mohanty}
\affiliation{Imperial College London, 1010 Blackett Lab, Prince Consort Road, London SW7 2AZ, UK}

\author{Giuseppe Lodato}
\affiliation{Dipartimento di Fisica, Universit\`a Degli Studi di Milano, Via Celoria, 16, Milano, I-20133, Italy}

\begin{abstract}
In this paper, we investigate the relation between disk mass and mass accretion rate to constrain the mechanism of angular momentum transport in protoplanetary disks.
We find a correlation between dust disk mass and mass accretion rate in Chamaeleon I with a slope that is close to linear, similar to the one recently identified in Lupus.
We investigate the effect of stellar mass and find that the intrinsic scatter around the best-fit \Mdust--\Mstar and \Macc--\Mstar relations is uncorrelated.
We simulate synthetic observations of an ensemble of evolving disks using a Monte Carlo approach,
and find that disks with a constant $\alpha$ viscosity
can fit the observed relations between dust mass, mass accretion rate, and stellar mass, but over-predict the strength of the correlation between disk mass and mass accretion rate when using standard initial conditions.
We find two possible solutions. In the first one, the observed scatter in \Mdust and \Macc is not primordial, but arises from additional physical processes or uncertainties in estimating the disk gas mass. Most likely grain growth and radial drift affect the observable dust mass, while variability on large time scales affects the mass accretion rates. In the second scenario, the observed scatter is primordial, but disks have not evolved substantially at the age of Lupus and Chamaeleon I due to a low viscosity or a large initial disk radius.
More accurate estimates of the disk mass and gas disk sizes in a large sample of protoplanetary disks, either through direct observations of the gas or spatially resolved multi-wavelength observations of the dust with \texttt{ALMA}, are needed to discriminate between both scenarios or to constrain alternative angular momentum transport mechanisms such as MHD disk winds.
\end{abstract}


\section{Introduction}
Gas-rich dusty disks around pre-main sequence stars are the sites of planet formation, hence their evolution and dispersal affect when and what types of planets can form. Observations have established that accretion of disk gas onto the star is a ubiquitous phenomenon  \citep[e.g.][]{2016ARA&A..54..135H} that appears to drive the early evolution of protoplanetary disks  \citep[e.g.][]{2014prpl.conf..475A}. Yet, the physical mechanism by which gas loses angular momentum and accretes is still hotly debated \citep[see, e.g.][for comprehensive reviews on the topic]{2011ARA&A..49..195A,2014prpl.conf..411T}.

The prevailing view has been that turbulence in disks transports angular momentum outward, enabling disk material to flow radially inward. 
The most common approach to parameterize the strength of turbulence is to assume the scaling relation between the viscosity, $\nu$, and the disk properties proposed by \cite{1973A&A....24..337S}:
\begin{equation}\label{eq:alpha}
\nu = \alpha c_s h,
\end{equation}
in the notation of \cite{1981ARA&A..19..137P} where $c_s$ is the sound speed, $h$ is the disk scale height, and $\alpha$ is a dimensionless parameter that represents the efficiency of angular momentum transport. 
When assuming $\alpha$ is independent of time and radius,
it is possible to construct models that describe the disk thermal structure {\it and} its evolution \citep[e.g.][]{1998Icar..132..100S,2011ARA&A..49..195A}. The simplicity of these constant $\alpha$ disk models has led to their widespread use both to predict the evolution and dispersal of protoplanetary disks \citep[e.g.][]{2006MNRAS.369..229A,2011MNRAS.412...13O}, dust evolution \citep[e.g.][]{2012A&A...539A.148B}, and to connect disk evolution to planet formation and composition \citep[e.g.][]{2009A&A...501.1139M,2017MNRAS.469.3910C}.
Another approach is to assume the turbulence-induced viscosity is time-independent and scales radially with a power-law, in which case self-similar solutions can be developed to analytically describe the disk evolution \citep[e.g.,][]{1974MNRAS.168..603L}.

A different approach is to compute the viscosity that arises from some turbulent process and then relate it to $\alpha$ through the framework discussed above. Magneto-Rotational Instability (MRI, \citealt{1991ApJ...376..214B}) is thought to be the leading mechanism to drive turbulence in disks while other instabilities
such as gravitational \citep[e.g.][]{2016ARA&A..54..271K} or hydrodynamic \citep[e.g.][]{2017arXiv170406786M}
are likely to play a minor role \citep[e.g.][]{2014prpl.conf..411T}. 
Global magneto-hydrodynamic (MHD) simulations of accretion disks in the ideal limit support this view and find a rate of angular momentum transport $\alpha$ of $10^{-3}-10^{-2}$ with modest radial variations \citep{2011ApJ...735..122F,2013A&A...560A..43F}.

However, it was long realized that MRI cannot operate in the entire disk, especially at intermediate radii ($\sim$1-10\,au) where the midplane is cool and shielded from ionizing radiation. This led to the development of layered disks in which accretion occurs primarily through an active ionized surface \citep{1996ApJ...457..355G}. In the dead zone, turbulent stress decreases by orders of magnitude \citep[e.g.][]{2017ApJ...835..230F} and the assumption that $\alpha$ is a global constant breaks down \citep[e.g.][]{2013ApJ...764...65M}. The inclusion of non-ideal MHD effects further complicates this picture as simulations suggest that accretion is shut off even in the disk surface \citep[e.g.][]{2013ApJ...769...76B,2013MNRAS.434.2295K,2015ApJ...801...84G} but strong winds develop that  extract angular momentum and enable disk accretion. If these winds dominate the angular momentum transport, the evolution of protoplanetary disks cannot be described by $\alpha$ disk models \citep[e.g.][]{2015ApJ...815..112K,2016ApJ...821...80B}.

Direct observational estimates of the turbulent motions of gas are only available for few disks \citep{2016A&A...592A..49T,2017ApJ...843..150F}.
In the context of $\alpha$ disk models, the observed disk masses, sizes, mass accretion rates and lifetimes suggest $\alpha$ of order $\sim0.01$ \citep{1998ApJ...507..361S,1998ApJ...495..385H,2010ApJ...723.1241A} or smaller \citep{2017ApJ...837..163R}.
However, the steep mass accretion rate--stellar mass relation (\Macc$\sim$\Mstar$^2$, e.g. \citealt{2006A&A...452..245N,2009A&A...504..461F,2014A&A...561A...2A}) remains challenging to explain. \cite{2006ApJ...648..484H} suggest that the steepness results from disks around very low-mass stars being less massive, fully magnetically active, and as such having viscously evolved substantially. On the opposite, \cite{2014MNRAS.439..256E} propose that the relation is caused by a specific disk dispersal mechanism, stellar X-ray driven photo-evaporation. Interestingly, \cite{2006ApJ...639L..83A} and \cite{2006ApJ...645L..69D} point out that the \Macc-\Mstar relation may not reflect disk evolution but rather the initial conditions of star formation, specifically the disk size.

\begin{table*}
\tiny
	\title{Stellar and disk properties for Lupus and Chamaeleon I.}
   	\begin{tabular}{cccccccccccccccc}\hline\hline
2MASS ID & Sp.T. & \Teff & $L_\star$ & $R_\star$ & \Mstar & err & $L_{\rm acc}$ & \Mdust & err & Detect & \Macc & err & Detect & region & Exclude \\
unit &  & K & [$L_\odot$] & [$R_\odot$] & [$M_\odot$] & [$M_\odot$] & [$L_\odot$] & [$M_\oplus$] & [$M_\oplus$] & T/F & [$M_\odot/$yr] & [$M_\odot/$yr] & T/F &  & \\
	\hline
J10533978-7712338    & M2 & 3560 & -1.80 & -0.48 & -0.41 & 0.11 & -4.56 & 0.19 & 0.0746 & True & -12.03 & 0.29 & True & Cha I & underlum\\
J10555973-7724399    & K7 & 4060 & -0.74 & -0.07 & -0.13 & 0.05 & -1.25 & 1.06 & 0.0168 & True & -8.58 & 0.28 & True & Cha I & \\
J10561638-7630530    & M6.5 & 2935 & -1.10 & 0.04 & -0.96 & 0.07 & -4.55 & 0.12 & 0.0174 & True & -10.95 & 0.28 & False & Cha I & \\
J10563044-7711393    & K7 & 4060 & -0.37 & 0.12 & -0.07 & 0.17 & -2.24 & 1.59 & 0.0041 & True & -9.45 & 0.32 & True & Cha I & \\
J10574219-7659356    & M3 & 3415 & -0.28 & 0.32 & -0.52 & 0.05 & -1.98 & 0.48 & 0.0395 & True & -8.54 & 0.27 & True & Cha I & \\
... 				  & ... & ... & ...   & ...  & ...   & ...  & ...   & ...  & ...    & ...  & ...   & ...  & ...  & ...   & ... \\ 
   	\hline\hline\end{tabular}
   	\caption{Stellar and disk properties for Chamaeleon I and Lupus. Columns 10 and 13 indicate if a source is detected in dust continuum emission ($>3\sigma$) and whether the accretion luminosity is larger than that expected from chromospheric emission. The last column denotes the reason why sources are not included. Table 1 is published in its entirety in the machine-readable format. A portion is shown here for guidance regarding its form and content
	}
	\label{t:data}
\end{table*}

The zeroth order expectation of viscously evolved disks
is that their mass accretion rate correlates linearly with disk mass \citep[e.g.][]{2006ApJ...645L..69D,2017MNRAS.468.1631R}.  Recent surveys of nearby star-forming regions are enabling for the first time to test this prediction on statistically significant samples where mass accretion rates and disk masses are available for the same objects. \cite{2016A&A...591L...3M} used 66 objects from the $\sim 1-3$\,Myr-old  Lupus star-forming region with mass accretion rates homogeneously  computed from VLT/X-Shooter spectra and disk masses from sub-mm continuum 
emission from the Atacama Large Millimeter/submillimeter Array (ALMA).
The relation between mass accretion rates and dust disk masses is found to be roughly consistent with viscously evolved disks.

Here, we expand upon this study by combining the \Alma and \Xshooter surveys of disks in the Lupus and Chamaeleon I star-forming regions, thus more than doubling the sample of \cite{2016A&A...591L...3M} (Section \ref{s:sample}).
First, we investigate the relation between dust mass, mass accretion rate, and stellar mass (Section \ref{s:analysis}). Then, we simulate a population of constant $\alpha$ disks using a Monte Carlo approach and quantify how the observed dust mass and mass accretion rate 
deviate from the simulated one (Section \ref{s:sim}).  Finally, we discuss the implications of our results and what observations/analysis should be carried out to further constrain the angular momentum transport in protoplanetary disks.
(Section \ref{s:discussion}).

\section{Homogeneous Analysis Of Stellar and Disk Properties}\label{s:sample}
We perform a homogeneous analysis of the dust disk mass, mass accretion rate, and stellar mass in the Chamaeleon I and Lupus star-forming regions. All observational data used in this analysis were previously published; the \Alma data surveys of disk masses were presented by \cite{2016ApJ...828...46A} and \cite{2016ApJ...831..125P}; the \Xshooter surveys of mass accretion rates were presented by \cite{2014A&A...561A...2A,2017A&A...600A..20A}, and \cite{2014A&A...568A..18M,2016A&A...585A.136M,2017arXiv170402842M}. The dust mass and mass accretion in Lupus were jointly analyzed by \cite{2016A&A...591L...3M}.

We use the stellar properties derived by \cite{2016ApJ...831..125P} using the \cite{2015A&A...577A..42B} and (non-magnetic) \cite{2016A&A...593A..99F} evolutionary tracks to achieve a homogeneous dataset for both star-forming regions, and recalculate mass accretion rates from the accretion luminosity.
We also use the dust masses for Lupus derived by \cite{2016ApJ...831..125P} for consistency.
All data used in this paper, including error bars and upper limits, are presented in Table \ref{t:data}.

\subsection{Chamaeleon I}\label{s:obs:cha}
The dataset of Chamaeleon I is based on the \Alma survey by \cite{2016ApJ...831..125P} and the \Xshooter survey presented by \cite{2016A&A...585A.136M,2017arXiv170402842M}. 

\begin{figure}
    \includegraphics[width=\linewidth]{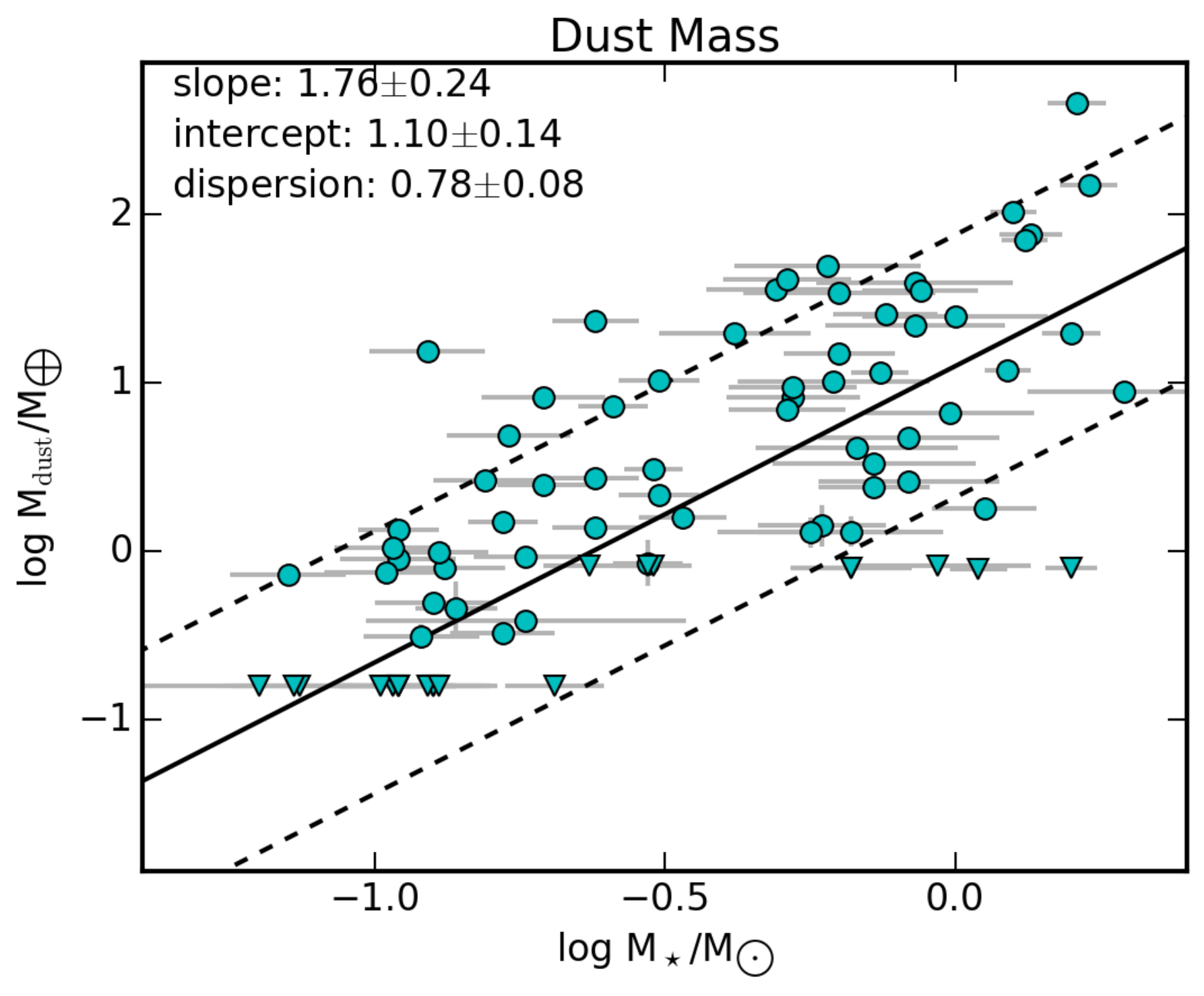}
    \caption{Dust mass ($\Tdust = 20 K$) versus stellar mass for sources in Chamaeleon I. Triangles denote 3$\sigma$ upper limits for sources that are not detected with \Alma. The solid line shows the best fit regression including upper limits. The $1\sigma$ dispersion around the best-fit is indicated with dashed lines. The \Mdust--\Mstar relation is steeper than linear consistent with previous work.}
    \label{f:mdust}
\end{figure}

Dust masses are taken from the \Alma continuum survey at $887 ~\mu$m from \cite{2016ApJ...831..125P}. Of the 93 sources, 66 are detected ($> 3 \sigma)$ and 27 have upper limits. 
We adopt the dust masses derived with a temperature of $T_{\rm dust}=20 K$.
Although the dust temperature may scale with stellar luminosity, and hence mass, \citep{2013ApJ...771..129A}, this assumption is dependent on how the disk outer radius scales with stellar mass \citep{2017ApJ...841..116H}.
Using a stellar-mass-independent temperature avoids introducing a correlated error between dust mass and stellar mass. 
While a disk temperature that decreases with stellar mass flattens the \Mdust--\Mstar relation \citep[e.g.][]{2016ApJ...831..125P} and weakens the \Mdust--\Macc relation \citep{2016A&A...591L...3M}, the intrinsic scatter around the \Mdust--\Macc relation remains unchanged. Hence, we focus our analysis on understanding the scatter more than the slope of the \Mdust--\Macc relation.
After removing stars without a mass accretion rate measurement from the \Xshooter survey (see below), the sample of stars discussed here has 63 detections with \Alma in dust continuum and 21 upper limits (Fig. \ref{f:mdust}).

Accretion luminosities were taken from the \Xshooter survey presented by \cite{2017arXiv170402842M}. 
Of the 93 sources targeted with X-shooter, 9 sources have no accretion measurement for reasons listed in the last column of Table \ref{t:data}. We do not remove known transition disks from the sample as they do not appear to be outliers based on their dust masses and mass accretion rates (see section \S \ref{s:m2}).
The sample consists of 67 accreting sources and 15 dubious accretors. Dubious accretors are stars with an emission line strength consistent with chromospheric activity, see \cite{2017arXiv170402842M} for details. We will display them as upper limits in all figures, and verify throughout the paper that treating them as upper limits or detections does not influence our results.

\begin{figure}
    \includegraphics[width=\linewidth]{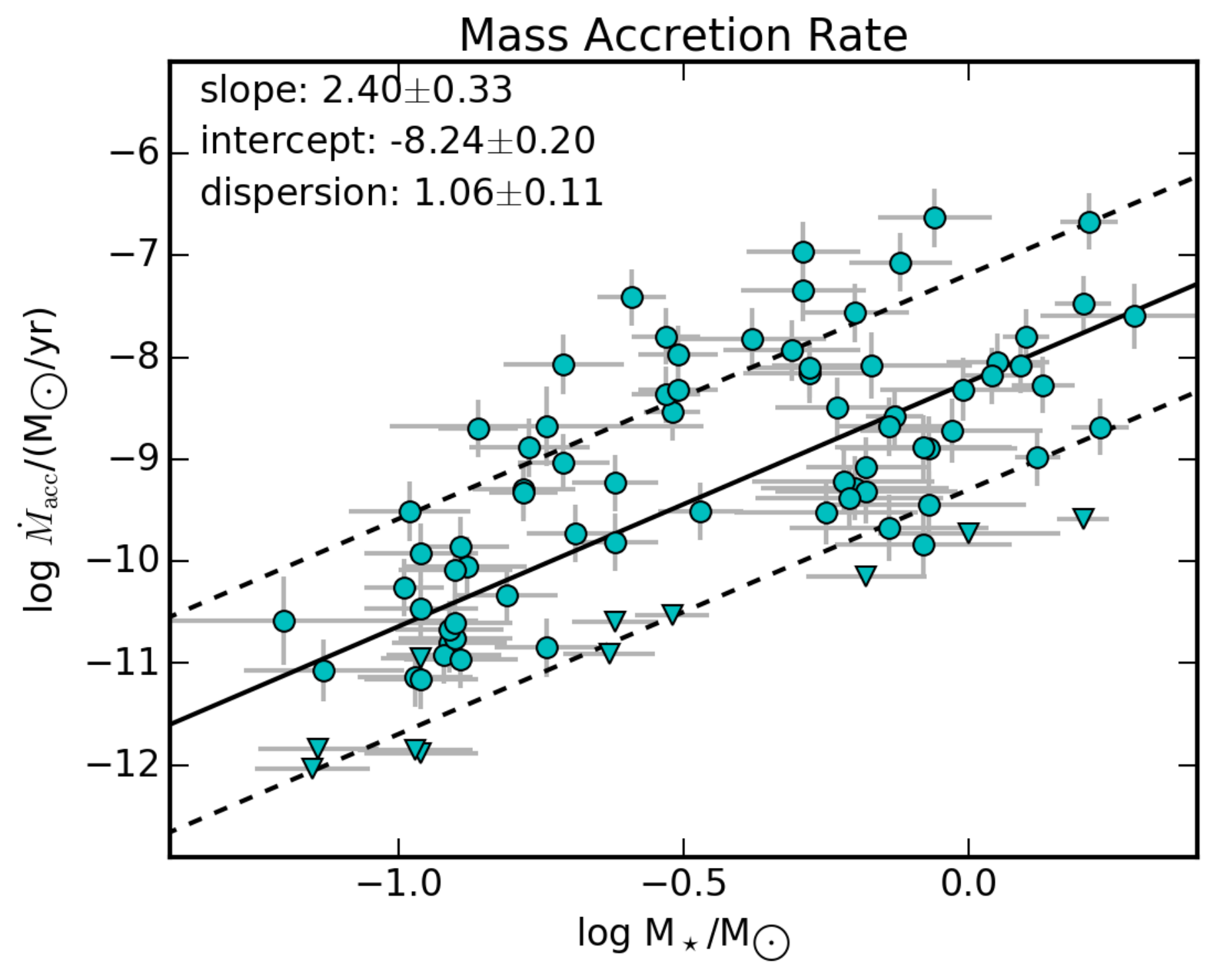}
    \caption{Mass Accretion rates versus stellar mass for sources in Chamaeleon I. Triangles denote dubious accretors for sources with an accretion luminosity consistent with chromospheric activity. The solid line shows the best fit regression treating dubious accretors as upper limits. The $1\sigma$ dispersion around the best-fit is indicated with dashed lines. The \Macc--\Mstar relation is quadratic consistent with previous work.}
    \label{f:macc}
\end{figure}

We calculate the mass accretion rate, $\dot{M}_{\rm acc}$, from the accretion luminosity, $L_{\rm acc}$, following \cite{1998ApJ...495..385H}:
\begin{equation}\label{eq:MaccLacc}
\dot{M}_{\rm acc}= 1.25 ~\frac{L_{\rm acc} R_\star}{G M_\star},
\end{equation}
where $R_\star$ is the stellar radius, \Mstar the stellar mass, and $G$ is the gravitational constant. The pre-factor 1.25 corresponds to a magnetospheric cavity size of 5 stellar radii, chosen to be consistent with \cite{2017A&A...600A..20A} and \cite{2017arXiv170402842M}. We propagate the errors on accretion luminosity ($0.25$ dex) and stellar parameters to calculate the error on the mass accretion rate, which is typically $0.3$ dex (see table \ref{t:data}). The difference in the accretion rates compared to \cite{2017arXiv170402842M} using the \cite{2000A&A...358..593S} and \cite{2015A&A...577A..42B} evolutionary tracks is small, with a median deviation of $\sim1\%$ and a maximum of $6\%$. The mass accretion rates as a function of stellar mass are shown in Figure \ref{f:macc}.

\subsection{Lupus}\label{s:obs:lup}
The dataset of Lupus is based on the \Alma survey by \cite{2016ApJ...828...46A} and the \Xshooter survey by \cite{2014A&A...561A...2A,2017A&A...600A..20A}. We follow the same procedure to derive a consistent dataset as for Chamaeleon I. For consistency, we use the stellar masses and dust masses from \cite{2016ApJ...831..125P}, which were derived using the same stellar evolutionary models and assumptions for the dust temperature and opacity as those for Chamaeleon I. We re-calculate the mass accretion rate from the accretion luminosity using Eq. \ref{eq:MaccLacc}. The difference in mass accretion rates with those derived using the \cite{2000A&A...358..593S} and \cite{2015A&A...577A..42B} evolutionary tracks by \cite{2017A&A...600A..20A} are again small, with a median deviation of $\sim4\%$ and a maximum of $30\%$.

\section{Analysis}\label{s:analysis}
We first analyze the correlation between dust mass, stellar mass, and mass accretion rate for Chamaeleon I. In \S \ref{s:combined} we present a joint analysis including the Lupus data. Throughout this section, we use the \texttt{Python} version of \texttt{linmix}\footnote{https://github.com/jmeyers314/linmix} \citep{2007ApJ...665.1489K} for linear regression to estimate best-fit parameters for the mean slope and intercept, the intrinsic dispersion around the mean trend, and the correlation coefficient. \texttt{Linmix} takes into account measurement errors in both dimensions and upper limits (censored data) in one dimension. 

\subsection{Chamaeleon I} 
The dependence of dust mass and mass accretion rate on stellar mass have previously been derived by \cite{2016ApJ...831..125P} and \cite{2017arXiv170402842M}, respectively. We re-fit these correlations to verify that our sample selection and the use of different stellar evolutionary models from \cite{2017arXiv170402842M} do not influence our results.

Figure \ref{f:mdust} shows the measured dust masses, \Mdust, as a function of stellar mass, \Mstar. The best-fit relation between dust mass and stellar mass is described by
\begin{equation}
\logMdust = 1.8(\pm 0.2) \logMstar + 1.1 (\pm 0.1),
\label{eq:mdustmstar:fit}
\end{equation}
and shown as the gray line in Figure \ref{f:mdust}. The $1\sigma$ dispersion is $0.8 \pm 0.1$ dex (gray dotted lines) and the correlation coefficient is $r=0.7 \pm 0.1$. These results are consistent with those in \cite{2016ApJ...831..125P} within the reported uncertainties. 
Figure \ref{f:macc} shows the stellar mass accretion rate, \Macc, as a function of stellar mass, \Mstar. The best-fit relation between mass accretion rate and stellar mass, treating dubious accretors as upper limits, is described by 
\begin{equation}
\logMacc = 2.4(\pm 0.3) \logMstar -8.3(\pm 0.2),
\label{eq:maccmstar:fit}
\end{equation}
with a dispersion of $1.1 \pm 0.1$ dex and correlation coefficient of $r=0.7 \pm 0.1$. These results are, within the uncertainties, consistent with the linear regression in \cite{2017arXiv170402842M}.

\begin{figure}
    \includegraphics[width=\linewidth]{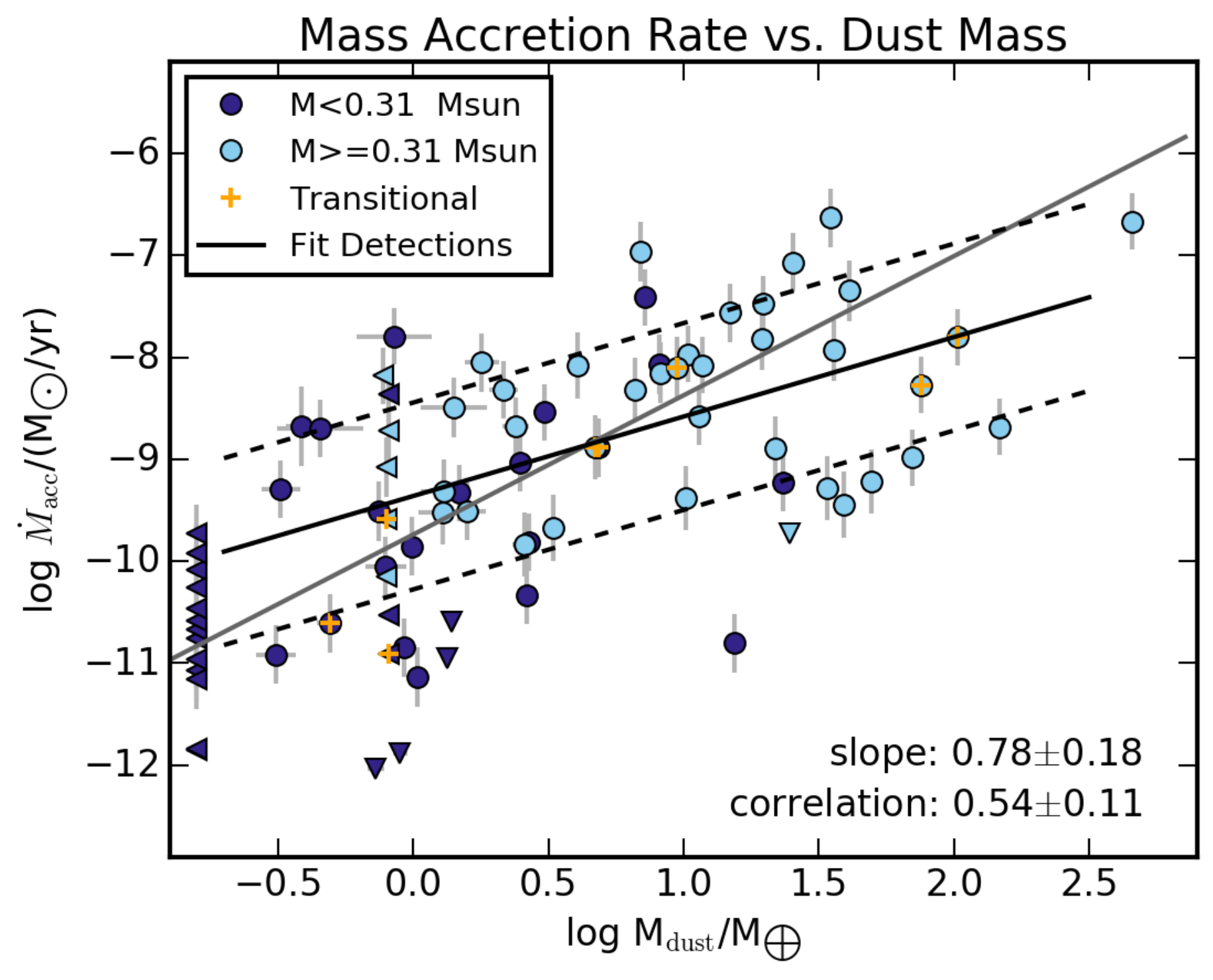}
    \caption{Dust masses ($\Tdust = 20 K$) versus mass accretion rates of sources in Chamaeleon I. Dubious accretors (left-facing triangles) and $3\sigma$ non-detections of the dust (down-facing triangles) are not included in estimating the best-fit \Macc--\Mdust relation (back solid line) and intrinsic dispersion (dashed lines) which has a correlation coefficient of $r\approx0.6$. Known transition disks are marked with yellow crosses and do not appear to be outliers. 
    The sample is color-coded by stellar mass, with the low-mass half in purple and high-mass half in light blue.
    The underlying distribution of stellar mass follows the expected correlation (gray solid line) based on the \Mdust--\Mstar and \Macc--\Mstar relations from Figures \ref{f:mdust} and \ref{f:macc}, respectively, with low(high) dust mass and low(high) mass accretion rates in bottom-left(top-right) corner.
    }
    \label{f:mdustmacc}
\end{figure}

\begin{figure}
    \includegraphics[width=\linewidth]{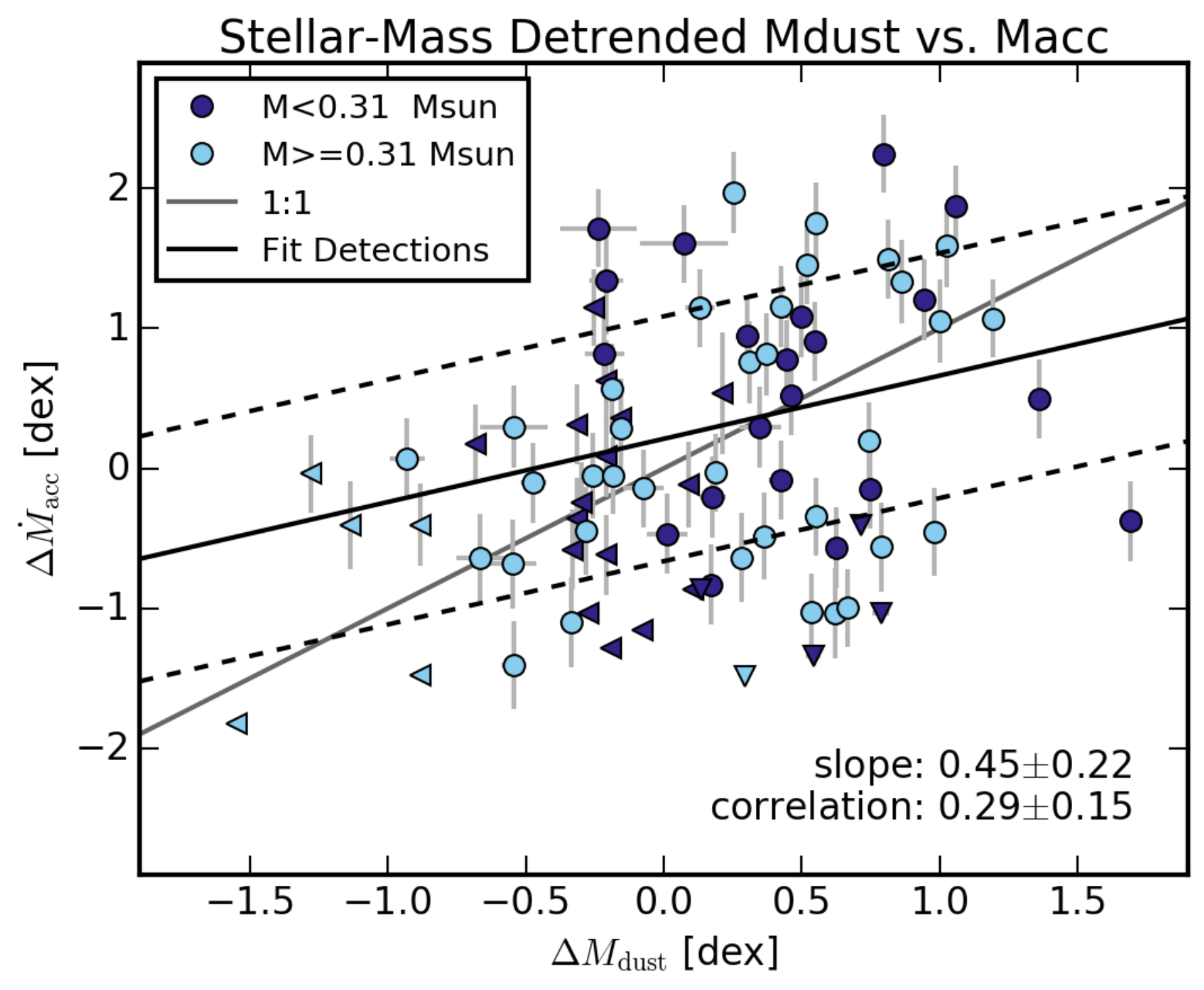}
    \caption{Stellar-mass-detrended mass accretion rate, \DA, versus dust mass, \DD. The gray line shows the linear correlation expected from disk models with a constant $\alpha$, which is not recovered by linear regression. There is no clear separation between the lower stellar mass (purple) and higher stellar mass (light blue) half of the sample. Black lines show the best-fit regression curve (solid line) and $1\sigma$ dispersion (dashed line). A correlation -- if present -- is weak ($r\approx0.2-0.4$, depending on how upper limits and dubious accretors are treated, see text). The lack of a clear correlation indicates that for stars of comparable mass, the mass accretion rate does not depend on dust mass. 
    }
    \label{f:mdustmaccmstar}
\end{figure}

\begin{figure*}
    \includegraphics[width=0.47\linewidth]{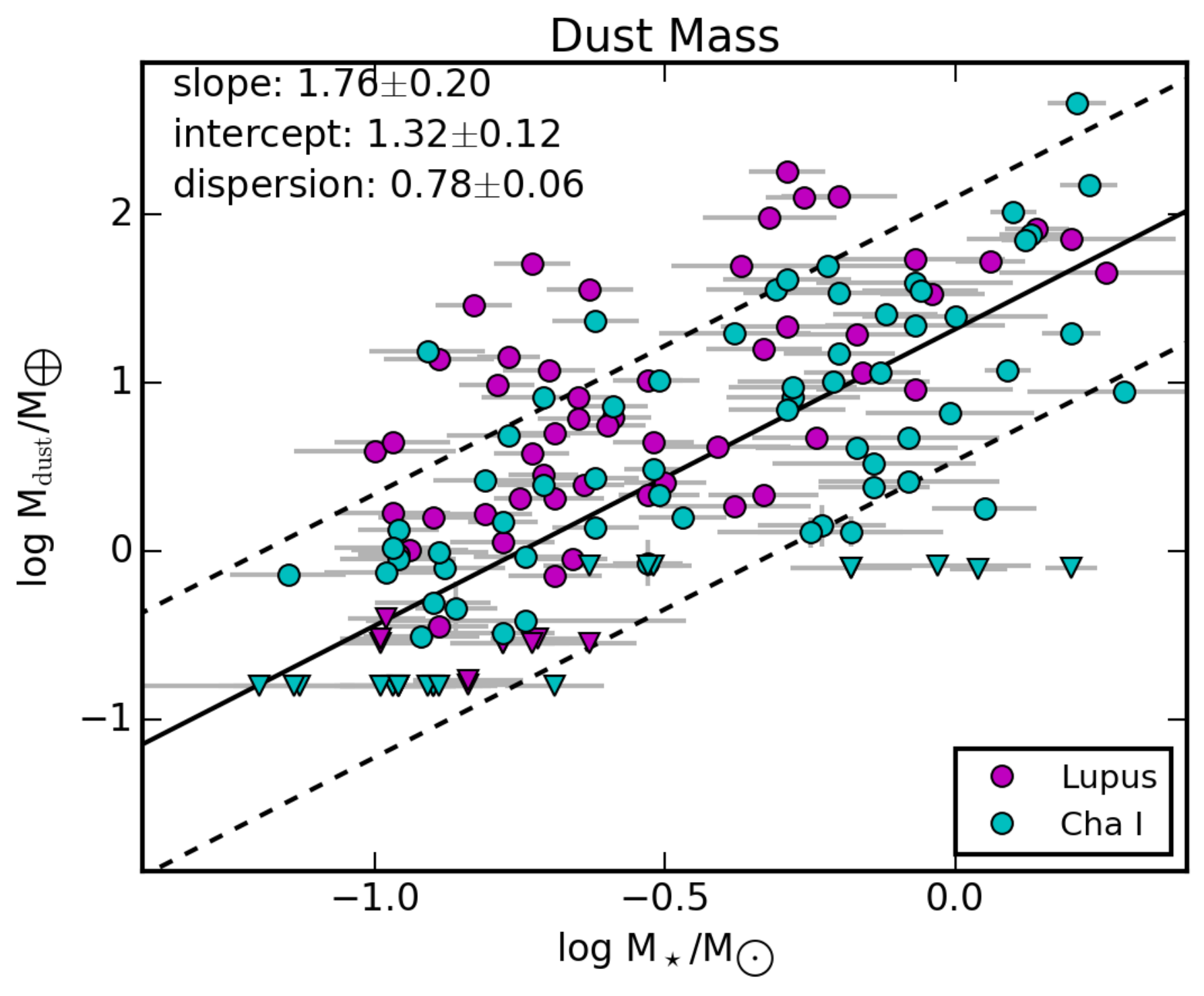}
    \includegraphics[width=0.47\linewidth]{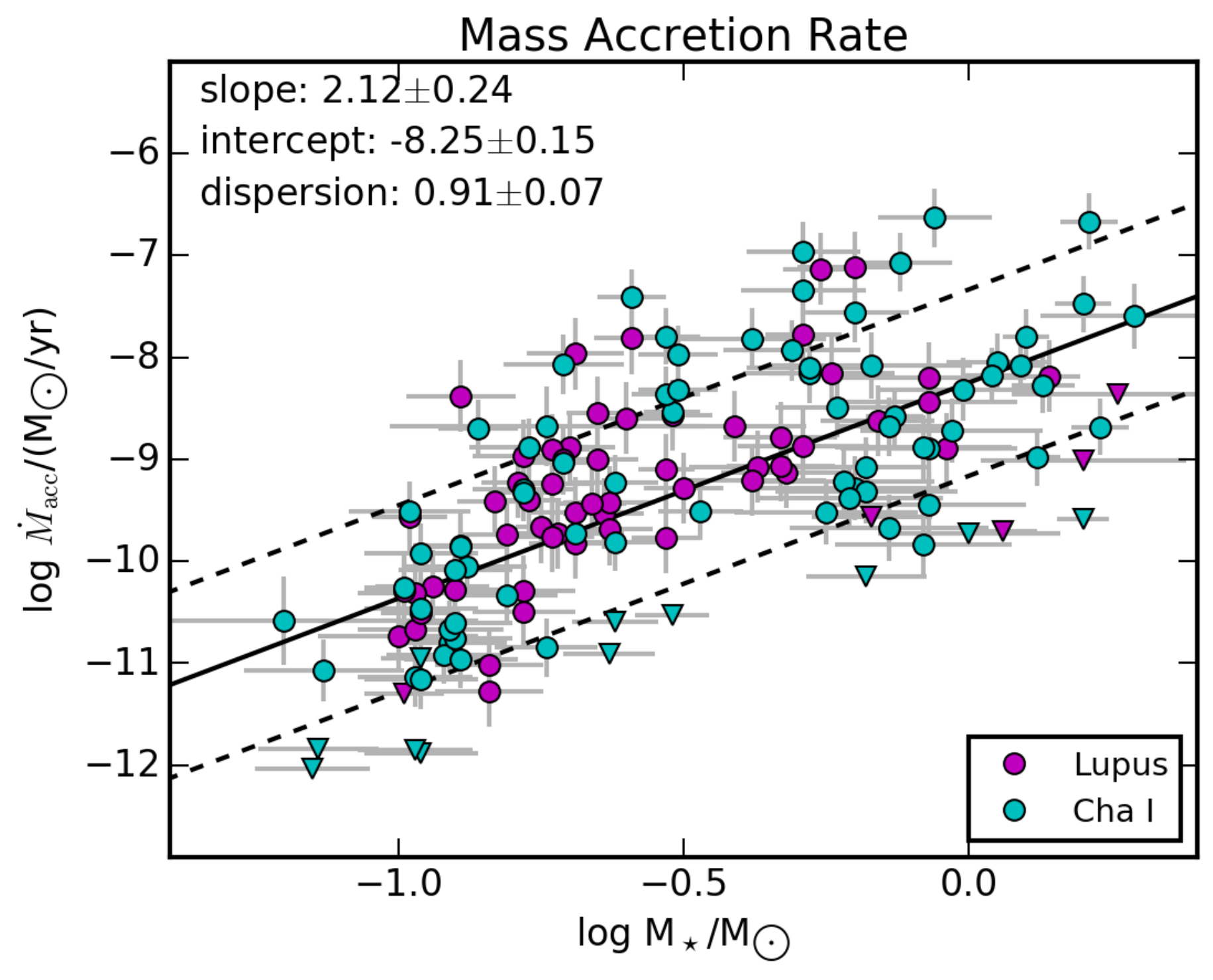}\\
	\includegraphics[width=0.47\linewidth]{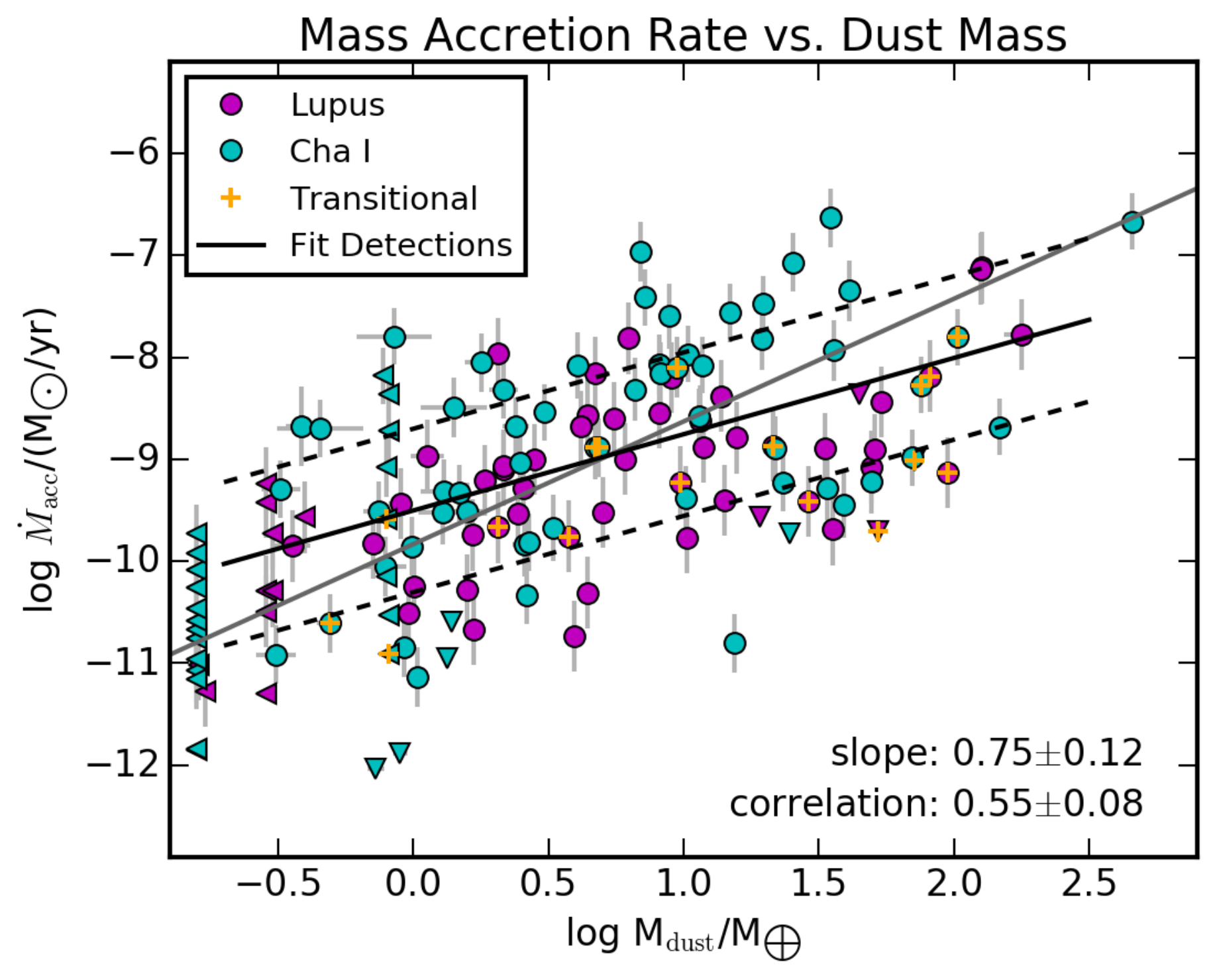}
	\includegraphics[width=0.47\linewidth]{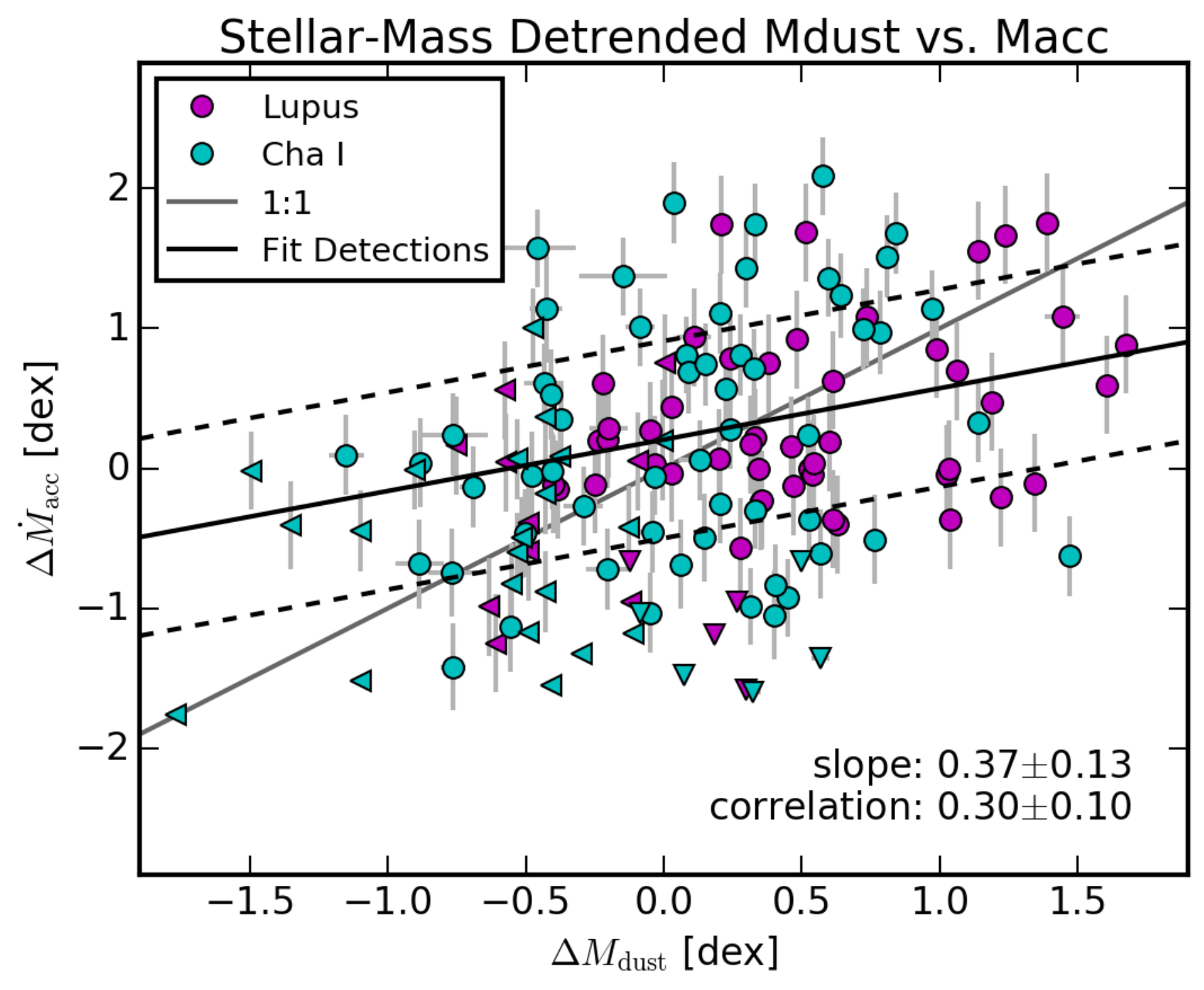}\\
    \caption{Dust masses, mass accretion rates, and stellar masses for the combined Chamaeleon I (cyan) and Lupus (pink) samples. Top left: Dust Mass versus stellar mass, as in Figure \ref{f:mdust}. Top right: Mass accretion rate versus stellar mass, as in Figure \ref{f:macc}. Bottom left: dust mass versus mass accretion rate as in Figure \ref{f:mdustmacc} but color-coded by star forming region. Bottom right: detrended dust mass versus mass accretion as in Figure \ref{f:mdustmaccmstar} but again color-coded by star forming region.
     }
    \label{f:lupcha}
\end{figure*}

\subsubsection{Mass Accretion Rate versus Dust Mass}\label{s:m2}
We investigate the relation between the dust mass and mass accretion rates following the analysis in \cite{2016A&A...591L...3M}. Figure \ref{f:mdustmacc} shows the stellar mass accretion rate versus dust mass in Chamaeleon I. Known transition disks are marked in red, but do not appear to be outliers in this distribution,
motivating our choice in \S \ref{s:obs:cha} to include them in the sample.
 
We find a moderate correlation between \Mdust and \Macc, $r=0.6\pm0.1$, fitting only sources with detections in both dimensions. The best-fit relation is given by
\begin{equation}
\logMacc= 0.8(\pm 0.2) \logMdust - 9.3(\pm 0.2),
\end{equation}
with a dispersion of $0.9 \pm 0.4$ dex. These values are consistent with those derived for Lupus \citep{2016A&A...591L...3M}, except for the dispersion which is significantly larger, owing to the larger dispersion in mass accretion rates in Chamaeleon I compared to Lupus \citep{2017arXiv170402842M}. The slope is within errors consistent with a linear correlation between dust mass and mass accretion rate. 

The slope is flatter than the expected correlation based on Eqs. (\ref{eq:mdustmstar:fit}) and (\ref{eq:maccmstar:fit}) (solid gray line, $\log\Macc\propto 1.4\pm0.3 \log\Mdust$).
A concern could be that the flatter slope may have been introduced by omitting upper limits in the fitting procedure. Because \texttt{linmix} does not support upper limits in two dimensions simultaneously we separately fit the upper limits in each dimension. Including upper limits on \Macc yields a linear slope of $1.0\pm0.2$ while including upper limits on \Mdust yields a steeper-than-linear slope of $1.6\pm0.2$. The latter is more consistent with the \Macc--\Mdust correlation based on the stellar-mass dependencies of both parameters (Eqs. \ref{eq:mdustmstar:fit} and \ref{eq:maccmstar:fit}).
Given the uncertainties in the derived values we conclude that there is no clear evidence that the \Macc--\Mdust relation deviates from a linear trend.

\subsubsection{Characterizing the Observed Scatter}\label{s:detrend}
A linear correlation between dust mass and mass accretion rate is consistent with the idea that protoplanetary disks evolve viscously \citep[e.g.][and references therein]{2016A&A...591L...3M}.
There is, however, significant scatter around the best-fit \Mdust--\Macc relation not predicted by constant $\alpha$ disk models \citep[e.g.][]{2006ApJ...645L..69D}. To characterize the intrinsic scatter in observed dust masses and mass accretion rates,
we divide out the fitted trend with stellar mass from the dust mass and mass accretion rate. We define two new quantities, \DD and \DA, that represent how much a given source deviates from the best-fit value at its stellar mass. The stellar-mass-detrended dust mass, \DD, is given by:
\begin{equation}\label{eq:dd}
\DD = \logMdust - (A_d\logMstar+B_d)
\end{equation}
where the coefficients $A_d=1.77$ and $B_d=12.6$ are taken from Equation $\ref{eq:mdustmstar:fit}$. Positive values of \DD  indicate a higher-than-average dust mass at that stellar mass, 
i.e. the source is above the mean trend in Figure \ref{f:mdust} (solid line). Negative values of \DD indicate a lower-than-average dust mass and the source is located below the best-fit trend in Figure \ref{f:mdust}.

The stellar-mass-detrended mass accretion rate, \DA, is given by:
\begin{equation}\label{eq:da}
\DA = \logMacc - (A_a\logMstar+B_a)
\end{equation}
where the coefficients $A_a=2.4$ and $B_d=5.5\times10^{-9}$ are taken from Equation $\ref{eq:maccmstar:fit}$.
The stellar-mass-detrended values for the dust mass (\DD) and mass accretion rate (\DA) are shown in Figure \ref{f:mdustmaccmstar}.

There is no clear trend visible between the detrended quantities \DD and \DA, in contrast to the \Macc--\Mdust plot. The lower-mass (purple) and higher-mass (cyan) half of the sample show a similar spatial distribution, indicating no residual trend with stellar mass. 
Fitting detections only, a weak ($r=0.28\pm0.15$) correlation may be present, with a slope of $0.36\pm0.14$ that deviates significantly from a linear correlation. Including upper limits on the mass accretion rate or dust mass yields a weaker ($r=0.19 \pm 0.14$) or stronger correlation ($r=0.35 \pm 0.12$), respectively, with similar slopes. Treating dubious accretors as detections does not significantly affect these results. In all cases, a strong correlation as may be expected from constant $\alpha$-disk models 
is not present. We test for the robustness of this result by increasing the sample size in \S \ref{s:combined}.

\subsection{Chamaeleon I and Lupus Combined}\label{s:combined}
The dust masses, mass accretion rates, and stellar masses of the combined Lupus/Chamaeleon I dataset are shown in Figure \ref{f:lupcha}, with a figure layout equivalent to Figures \ref{f:mdust}--\ref{f:mdustmaccmstar}. 
The dust disk masses of both regions show a similar dependence on stellar mass, but one that is different from the older Upper Sco star forming region \citep{2016ApJ...831..125P}. The mass accretion rates show a similar mean trend with stellar mass (top right panel, see also \citealt{2017A&A...600A..20A,2017arXiv170402842M}), though the dispersion around the mean trend in Lupus is smaller \citep{2014A&A...561A...2A,2017A&A...600A..20A}. As discussed in \S \ref{s:m2}, the mass accretion rates show the same correlation with dust mass in Chamaeleon I as in Lupus. 

We recover the \Macc--\Mdust correlation in the combined Lupus-Chamaeleon I dataset with a correlation coefficient of $0.55\pm0.08$, a slope of $0.75\pm0.12$, and a dispersion of $0.80 \pm 0.07$ (bottom left panel of Figure \ref{f:lupcha}). These results are consistent with estimates for the individual star forming regions, but derived at higher statistical confidence due to the larger sample size. We show in Appendix \ref{s:limited} that considering a limited stellar mass range does not lead to a stronger correlation. We investigate a possible underlying correlation with stellar mass by detrending \Mdust and \Macc with the same procedure as described in \S \ref{s:detrend}. The stellar-mass detrended dust mass, \DD, is calculate from Eq. \ref{eq:mdustmstar:fit} with coefficients $A_d=1.75$ and $B_d=10^{1.3}$ derived from fitting the combined dataset (solid line in top left panel). The stellar-mass detrended mass accretion rate, \DA, is derived using Eq. \ref{eq:maccmstar:fit} with coefficients $A_a=2.1$ and $B_a=10^{-8.2}$ derived from fitting the combined dataset (solid line in top right panel).

We again find a weak correlation ($r=0.27\pm0.10$) between \DD and \DA, with a slope of $0.36\pm0.14$ that is inconsistent with linear. The inclusion of upper limits in either dimension and treating dubious accretors as upper limits do not significantly change the strength of the correlation.

\begin{table*}
\centering
	\title{Model Parameters}
   	\begin{tabular}{llllllllllllll}\hline\hline
	Model &   & \multicolumn{2}{l}{$M_{\rm disk,0}$(\Msun)} & \multicolumn{2}{l}{$R_{\rm out,0}$(au)} & \multicolumn{2}{l}{$t_{\rm disk}$(Myr)}	 & \multicolumn{2}{l}{$\alpha$} & \multicolumn{2}{l}{\fgtd} & \multicolumn{2}{l}{\facc} \\
	\hline
    \visa & Mean      &  $0.1 M_\star^{1.9}$ && 33 && 2 && 0.01 && 300 && 1 \\
          & \hspace{0.5cm} Disp.(dex) &&  0.8                 && 0.3 && 0.3 && 0.3 && 0 && 0 \\
          \hline
	\visb & Mean  	   &  $0.1 M_\star^{1.9}$ && 33 && 2 && 0.01 && 300 && 1 \\
          & \hspace{0.5cm} Disp.(dex) &&  0.3                 && 0.3 && 0.3 && 0.3 &&  0.4 && 0.7 \\
          \hline
    \visc & Mean  	   &  $0.4 M_\star^{1.9}$ && 33  && 2 && 0.001 && 1000 && 1 \\
          & \hspace{0.5cm} Disp.(dex) &&   0.5                && 0.5 && 0.5 && 2.0 &&  0 && 0 \\   
   	\hline\hline\end{tabular}
   	\caption{Initial conditions for the simulated disk models. Parameters for each disk are randomly drawn from a log-normal distribution with the listed mean and standard deviation (dispersion). \fgtd is the gas-to-dust ratio in the disk at time of observation. \facc is the ratio of the measured mass accretion rate to the time-averaged mass accretion rate, representing accretion variability.
	}
	\label{t:model}
\end{table*}

\section{Simulations}\label{s:sim}
In this section we make a quantitative comparison between disk evolutionary models and the observed relations between dust mass, mass accretion rate, and stellar mass. We take a forward-modeling approach, simulating an ensemble of evolving disks and generating synthetic observations that are analyzed with the same statistical tools and procedures as the observations.

We use the \cite{2009ApJ...705.1206C} analytic disk model to simulate the time-evolution of a protoplanetary disk. This model calculates the surface-density evolution of an irradiated disk due to a (turbulent) viscosity, parameterized by the dimensionless quantity $\alpha$ \citep{1973A&A....24..337S} (see Eq. \ref{eq:alpha} and introduction) which is kept constant throughout the disk and in time\footnote{Note that, unlike in self-similar solutions, viscosity changes in time and it is not restricted to a radial power-law dependence.}.

The disk evolution depends on a number of initial parameters. The stellar radius, $R_\star$, and luminosity, $L_\star$, are calculated from the stellar mass, \Mstar, using the combined (non-magnetic) \cite{2015A&A...577A..42B} and \cite{2016A&A...593A..99F} evolutionary tracks as in \cite{2016ApJ...831..125P}.  The initial disk mass ($M_{\rm disk,0}$), radius ($R_{\rm out,0}$), opacity, and viscosity are adopted from the first example in \cite{2009ApJ...705.1206C}, except for the last model which explores non-standard initial conditions (see Sect.~\ref{s:vis:3}). Each disk is evolved until the age of the system, $t_{\rm disk}$. A stellar-mass dependency of $\Mstar^{1.9}$ is introduced to the initial disk mass to fit the observed scalings between \Mdust-\Mstar. Because in a constant $\alpha$ disk the dust mass and mass accretion rate are coupled, this also introduces a stellar-mass dependency in the resulting \Macc--\Mstar relation. 

To match the observed scatter in \Mdust and \Macc, we introduce a dispersion in disk model parameters ($M_{\rm disk,0}$,$R_{\rm out,0}$, $t_{\rm disk}$, $\alpha$). The dispersion in initial disk mass and radius represent variations in disk initial conditions. The dispersion in disk life time represent an age spread in the cloud. The dispersion in viscosity-parameter $\alpha$ represent variations in angular momentum transport efficiency between disks. 
We also introduce two additional parameters that can contribute to the observed scatter. The gas-to-dust ratio, $\fgtd= M_{\rm gas}/\Mdust$, to convert the modeled gas disk mass to a dust mass. The dispersion in \fgtd reflects both physical process that may alter the gas-to-dust ratio (see \S \ref{s:otherprocess}) as well as uncertainties in deriving the dust mass from the unresolved millimeter flux (see \S \ref{disc:var}). The other parameter, $\facc=\Macc/\dot{M}_{\rm disk}$, represents accretion rate variability and is defined as the ratio of the observed instantaneous stellar mass accretion rate to the time-averaged mass accretion rate of the disk. 

We simulate a survey similar in size to the combined Lupus and Chamaeleon I sample, with 140 stars randomly drawn 
between $0.1$ and $1.6 \Msun$ from a log-normal initial mass function \citep{2003ApJ...586L.133C}. 
For each star we run the \cite{2009ApJ...705.1206C} disk model with free parameters ($M_{\rm disk,0}$, $R_{\rm out,0}$, $t_{\rm disk}$, $\alpha$, $\fgtd$, and $\facc$) randomly sampled from a log-normal distribution, with mean and standard deviation as in Table \ref{t:model}. 
The gas-to-dust ratio (\fgtd) was increased to 300 to fit the intercept of both \Mdust and \Macc. The choice of gas-to-dust ratio is not unique, and a ratio of 100 can be achieved by a different conversion of millimeter flux to dust mass, either by lowering the dust opacity or decreasing the dust temperature.
These values reproduce the best-fit \Mdust-\Mstar and \Macc-\Mstar relations for the combined Lupus-Chamaeleon I dataset (e.g. Figure \ref{f:lupcha}).

Synthetic observations are conducted using a Monte Carlo simulation by perturbing each observable (\Mstar, \Mdust, \Macc) with an observational uncertainty of $0.1$, $0.1$, and $0.25$ dex, respectively. 
The simulated observables are considered upper limits (or dubious accretors) if the dust mass (or mass accretion rate) falls below the detection threshold of the survey (see \citealt{2016ApJ...831..125P,2017arXiv170402842M} for details).

\begin{figure*}
   	\includegraphics[width=0.47\linewidth]{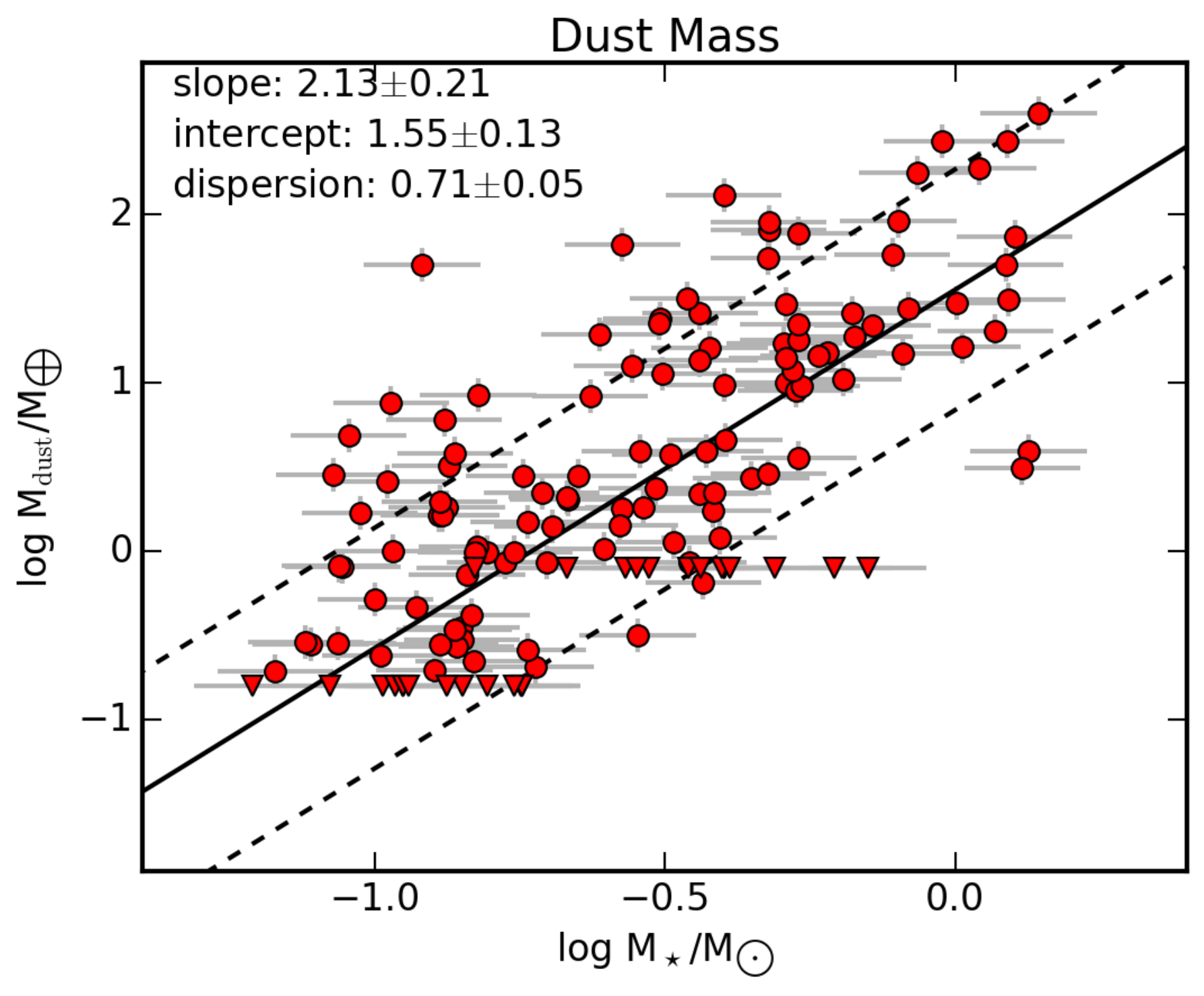}
   	\includegraphics[width=0.47\linewidth]{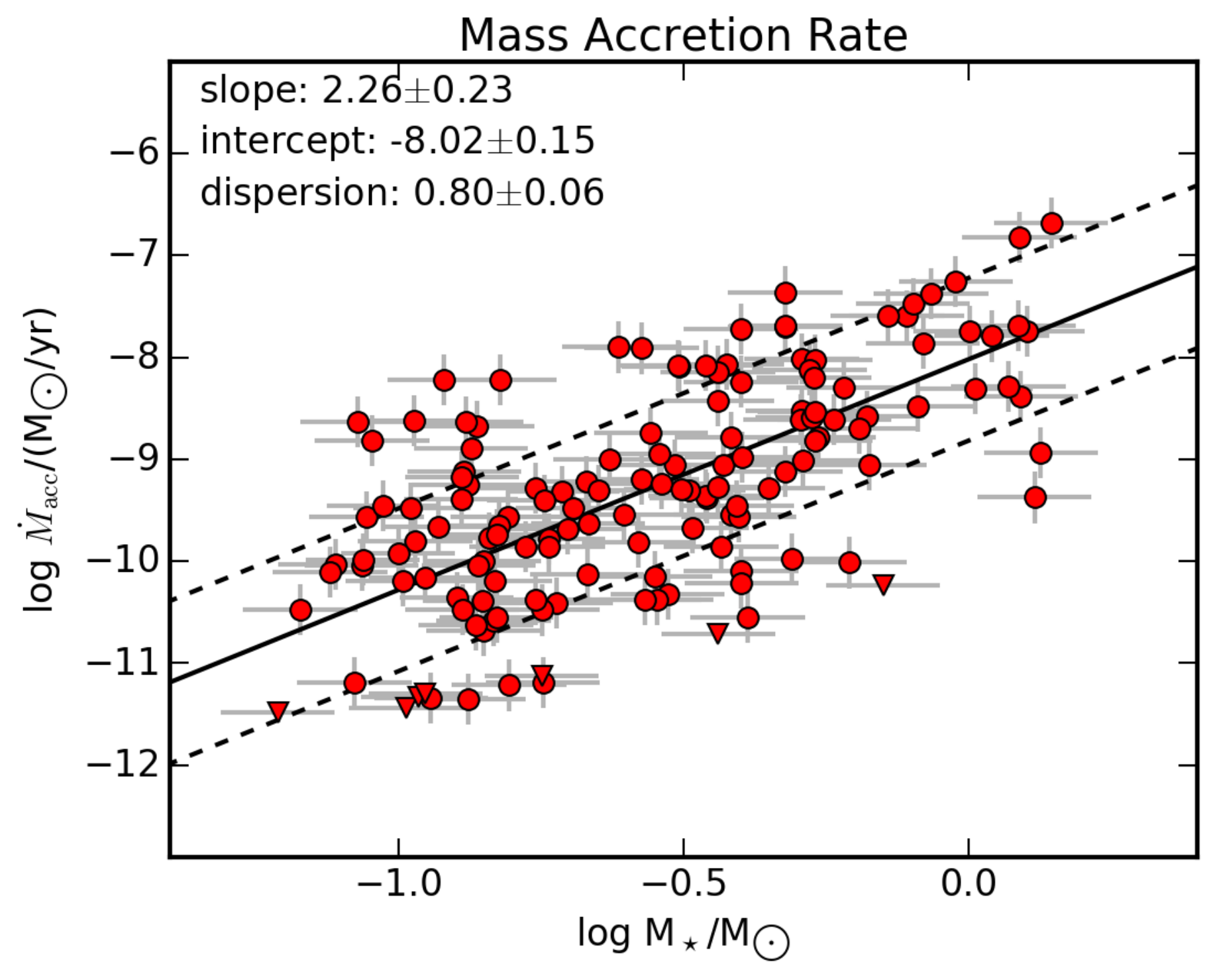}\\
	\includegraphics[width=0.47\linewidth]{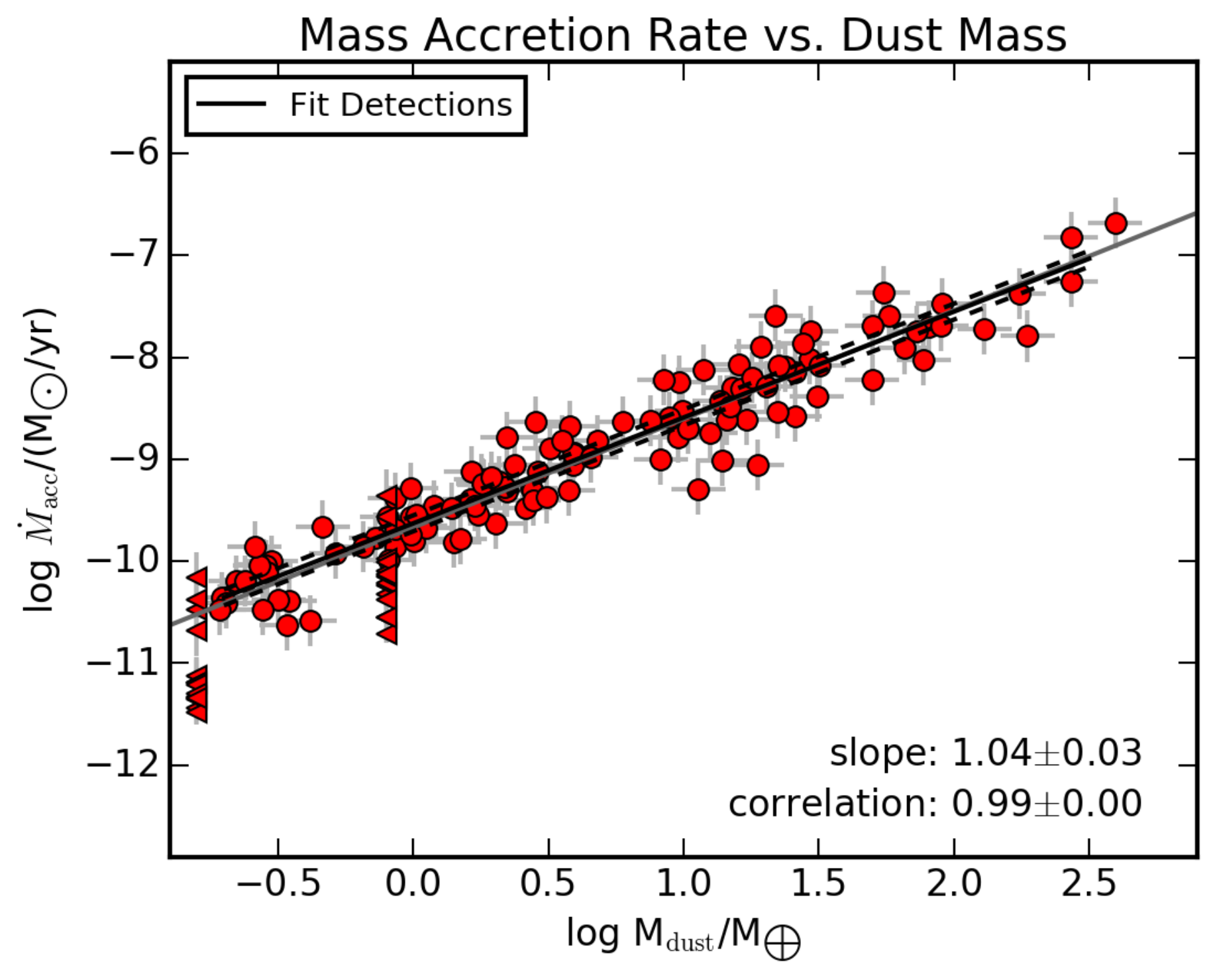}
	\includegraphics[width=0.47\linewidth]{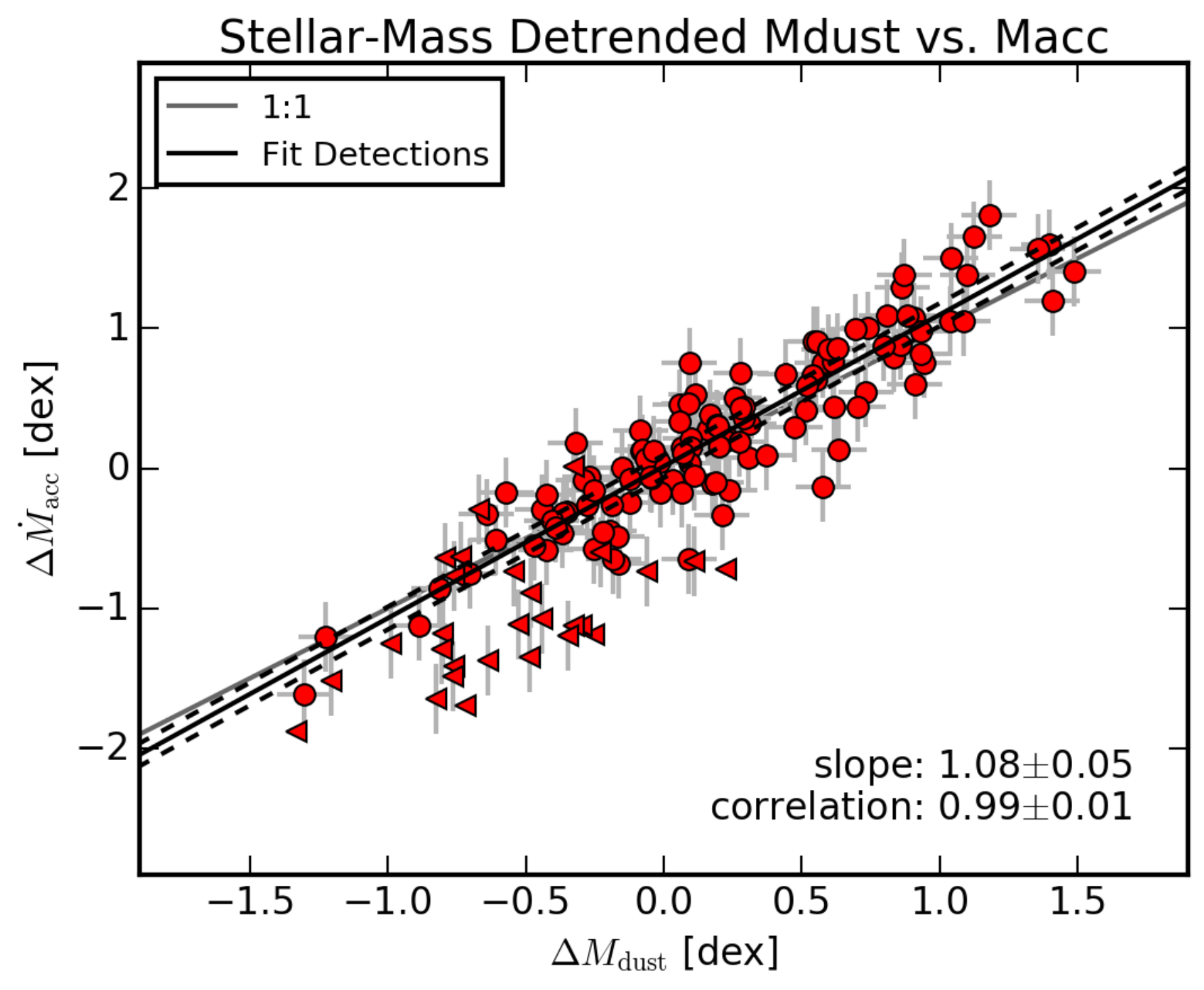}\\
    \caption{Synthetic observations for the disk model \visa described in \S \ref{s:vis:1}. The panel layout is the same as for the observed Lupus-Chamaeleon I dataset in Figure \ref{f:lupcha}. The inferred correlations in the bottom two panels are much stronger than observed, indicating that a dispersion in initial disk mass is not the main factor determining the dispersion in dust mass and mass accretion rates.
    }
    \label{f:vis:1}
\end{figure*}

\subsection{Model \visa}\label{s:vis:1}
First we simulate a disk model where the observed scatter in \Mdust and \Macc arises from a dispersion in initial disk conditions as in \cite{1998ApJ...495..385H,2006ApJ...645L..69D}.
The dust mass is assumed to be a direct tracer of the gas mass (no dispersion in $\fgtd$) and the instantaneous mass accretion rate is a direct tracers of the time-averaged mass accretion rate (no dispersion in $\facc$).
The stellar-mass dependencies in dust mass and mass accretion rate are, within their uncertainties, consistent with the observed values (Figure \ref{f:vis:1}).
Disk-to-disk variations in the initial disk mass, radius, $\alpha$, and age create a scatter of $\sim0.8$ dex around the best-fit \Mdust-\Mstar and \Macc-\Mstar relations. 

The \visa set of disk models has a median viscous time scale of 0.1 Myr, significantly shorter than the disk life time, and produce a nearly linear \Mdust--\Macc relation. Taking into account measurements errors and upper limits, the correlation is recovered at high confidence with negligible scatter. The correlation is much stronger than observed, with a correlation coefficient of unity compared to $r=0.6$ for the Lupus-Chamaeleon~I data. 

We use Eqs. \ref{eq:dd} and \ref{eq:da} to detrend the synthetic observations and calculate \DD and \DA, where the coefficients $A_D=1.4$ and $B_D=8.5$ are fitted to synthetic \Mdust-\Mstar observations and $A_A=2.0$ and $A_D=5\times 10^{-8}$ are fitted to synthetic \Macc-\Mstar observations. The strong linear correlation between \DD and \DA remains present in the simulated data after detrending.

\begin{figure*}
   	\includegraphics[width=0.47\linewidth]{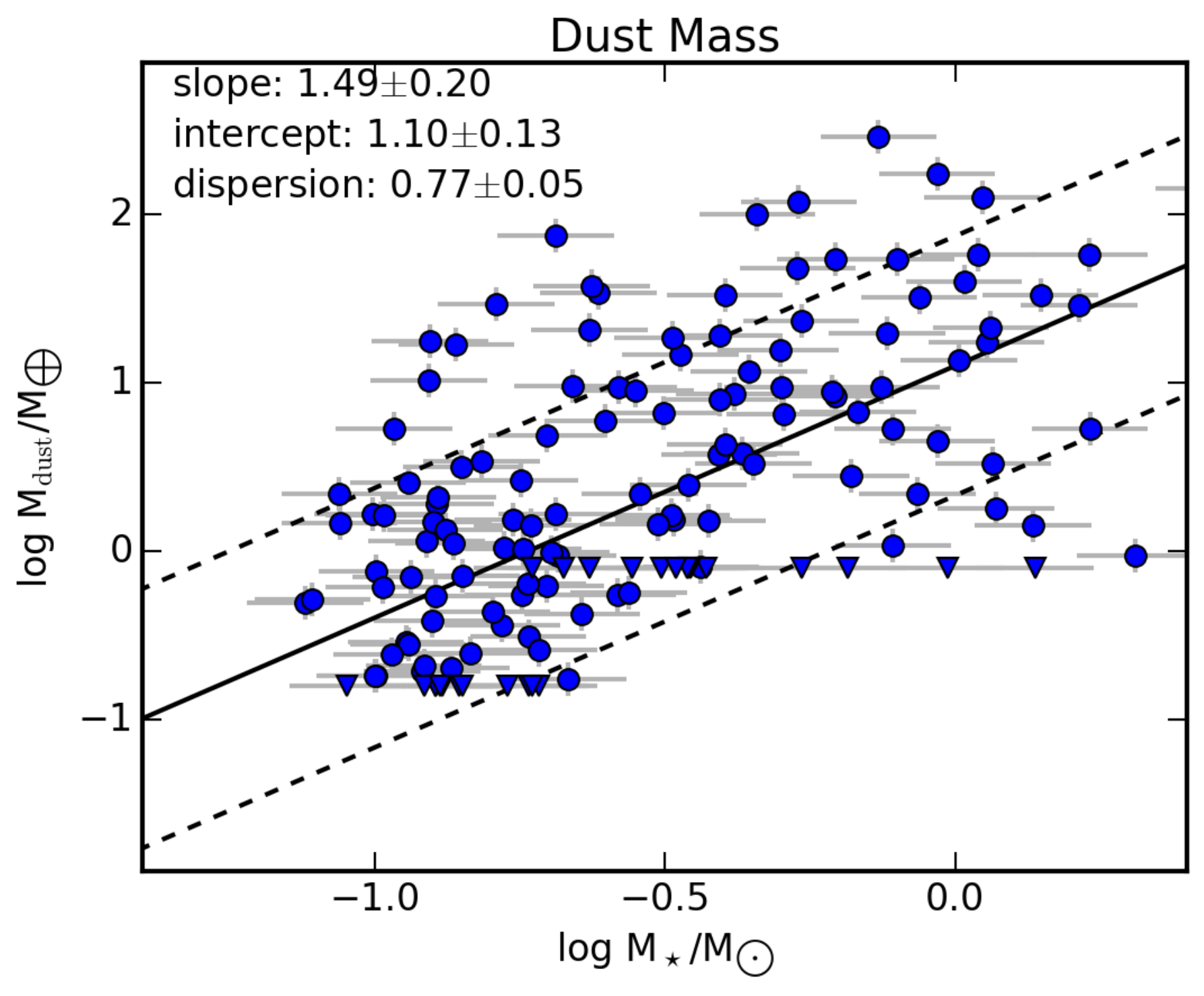}
   	\includegraphics[width=0.47\linewidth]{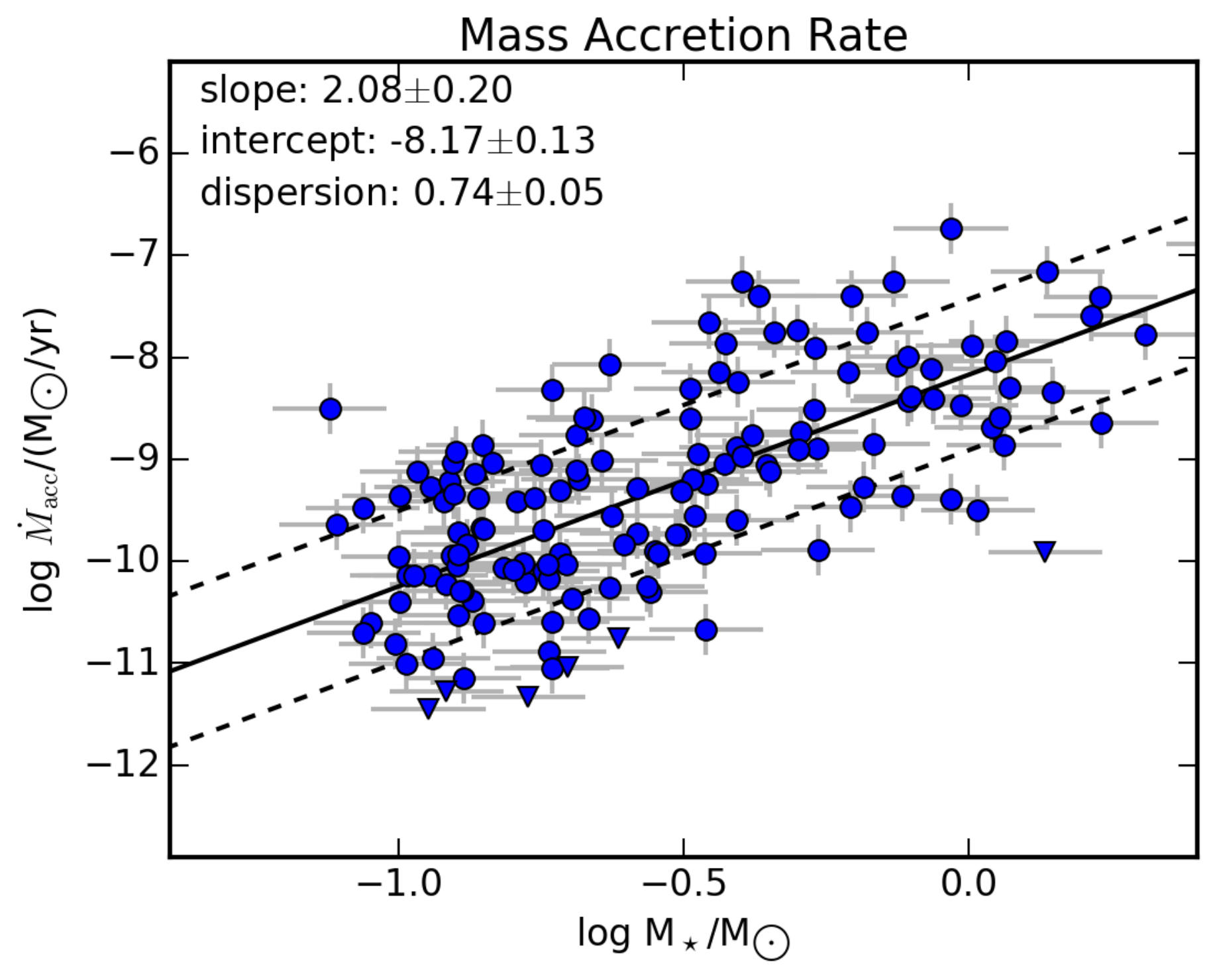}\\
	\includegraphics[width=0.47\linewidth]{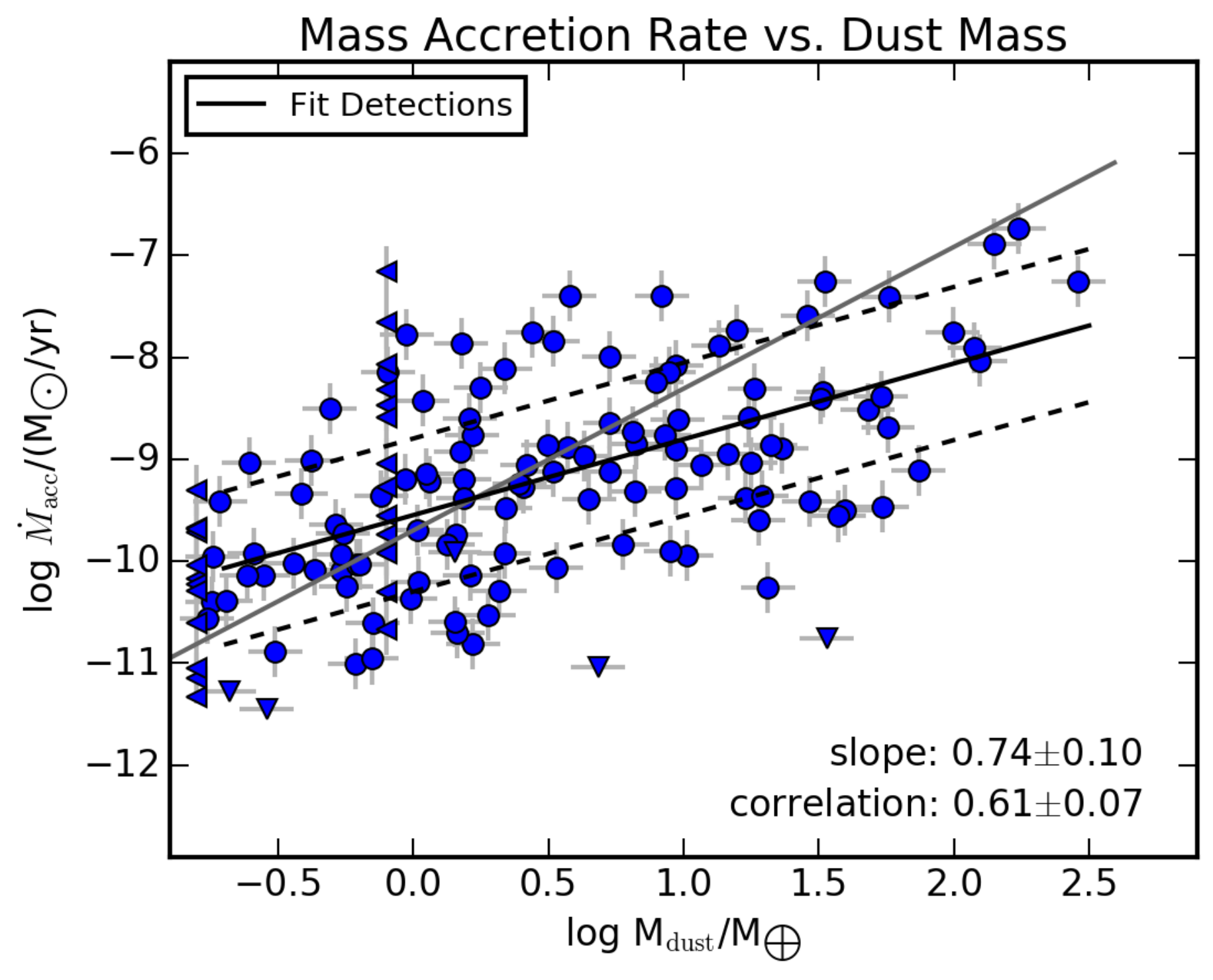}
	\includegraphics[width=0.47\linewidth]{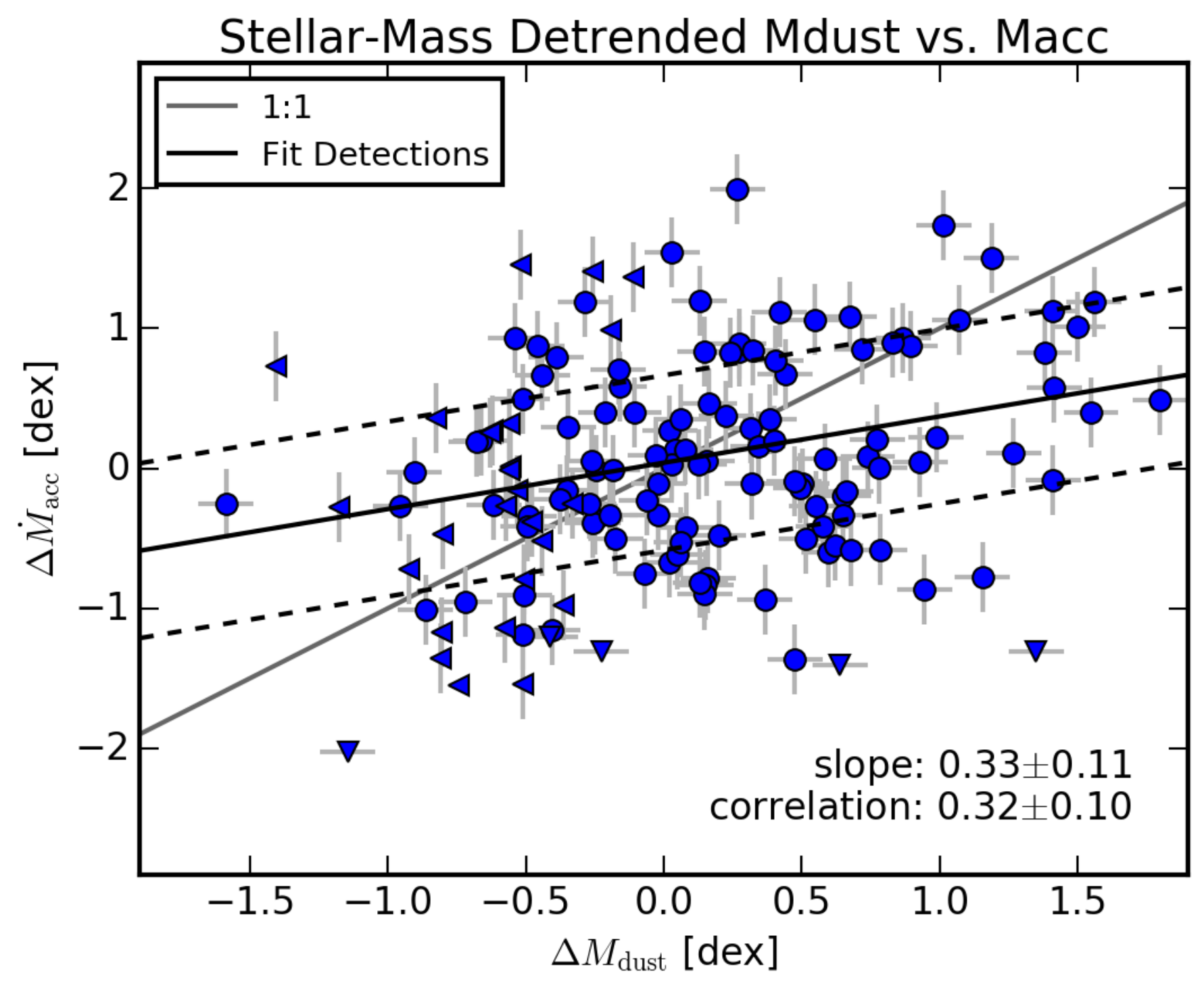}\\
      \caption{Synthetic observations for the disk model \visb described in \S \ref{s:vis:2}. Same panels as Figure \ref{f:vis:1}. 
    }
    \label{f:vis:2}
\end{figure*}

The \visa model presented here is not a unique solution. 
A degenerate set of parameters exist where the dispersion in initial disk mass can be traded off for higher dispersions in disk outer radius, life time, and/or viscosity, and vice versa without affecting the strength of the observed correlations.
The observable dispersion is less sensitive to these parameters than to the initial disk mass. For example, a dispersion in the outer disk radii of $0.8$ dex results in an scatter in the observed dust mass and mass accretion rates of $\sim0.3-0.4$ dex.
A dispersion in the disk mass of $0.5$ dex was independently derived by \cite{2003MNRAS.342.1139A} by modeling the fraction of stars with disks as function of time. 
We explored a large range of initial conditions (a factor 10 in initial disk mass and outer radius, a factor 100 in $\alpha$) and consistently find that these solutions produce strong correlations between \Mdust and \Macc ($r>0.95$) and \DD and \DA ($r>0.9$), except when long viscous timescales are used in combinations with non-standard input parameters (see Sect.~\ref{s:vis:3}, \visc model).
The strong correlations show that the zeroth-order assumption of dust mass and mass accretion rate as direct tracers of disk evolution are inconsistent with the observed moderate and weak correlations within the framework of a viscously evolved constant $\alpha$-disk model.

\subsection{\visb model}\label{s:vis:2}
The observed dust mass and mass accretion rate may not be perfect tracers of the disk conditions. 
Spatial and temporal variations in disk viscosity, as well as the accretion process near the stellar magnetosphere, lead to accretion rate variability. At the same time, variations from disk to disk in dust temperature, opacity, and gas-to-dust mass ratio may also contribute to the observed scatter in millimeter fluxes, hence dust masses.

Here, we explore how large the influence of these two processes needs to be for the constant $\alpha$ disk model to produce the observed scatter in the \Mdust--\Macc relation, modeled by parameters $\facc$ and $\fgtd$.
In model \visb we reduce the dispersion in the initial disk mass to $0.3$ dex, reducing the scatter in \Mdust and \Macc. We add scatter to the observable mass accretion rate by introducing variability in the accretion rate, \facc, of $0.7$ dex. Similarly, we increase the scatter in the observable dust masses by adding disk-to-disk variations in the gas-to-dust ratio, \fgtd, of $0.7$ dex. 

\begin{figure*}
   	\includegraphics[width=0.47\linewidth]{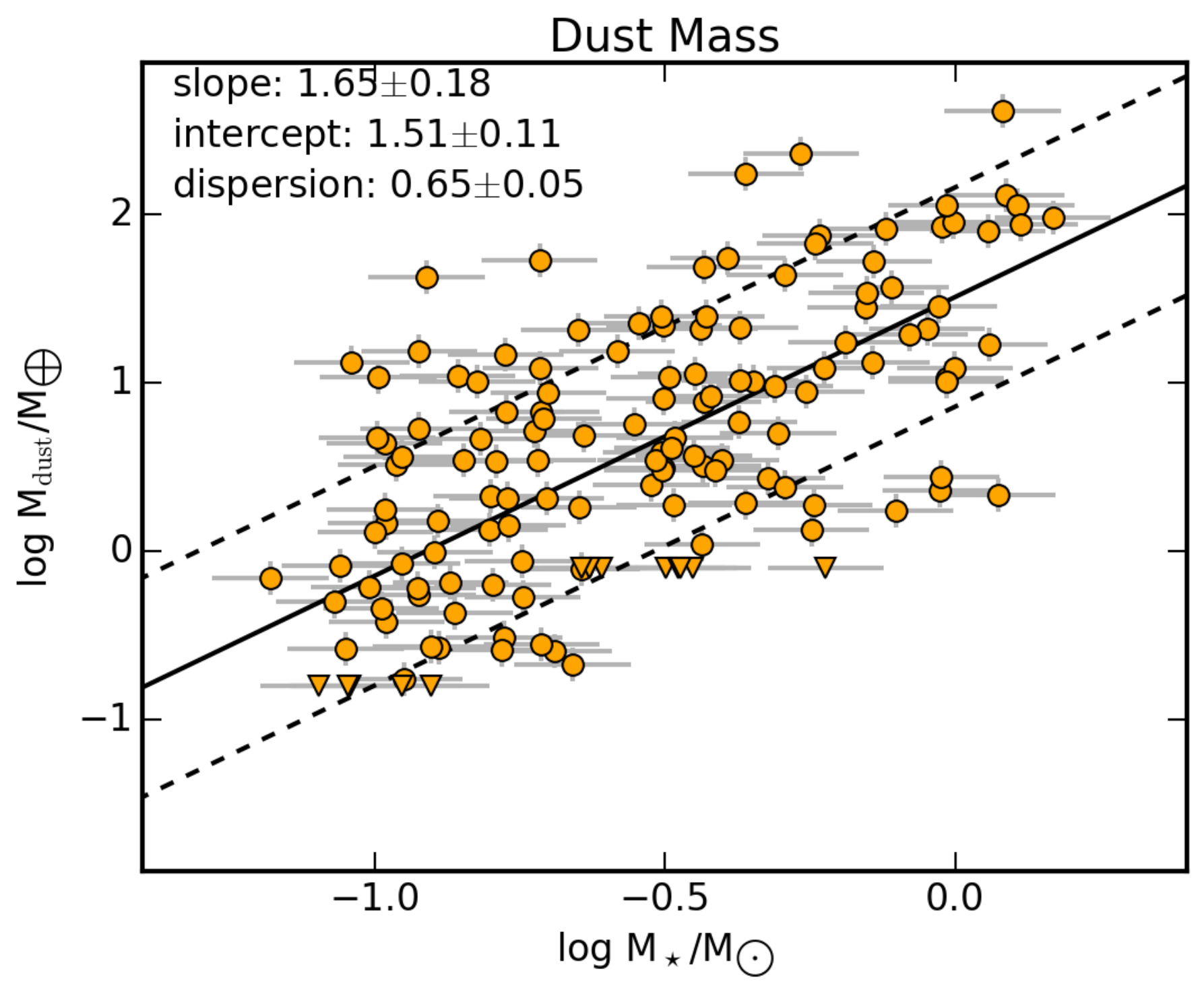}
   	\includegraphics[width=0.47\linewidth]{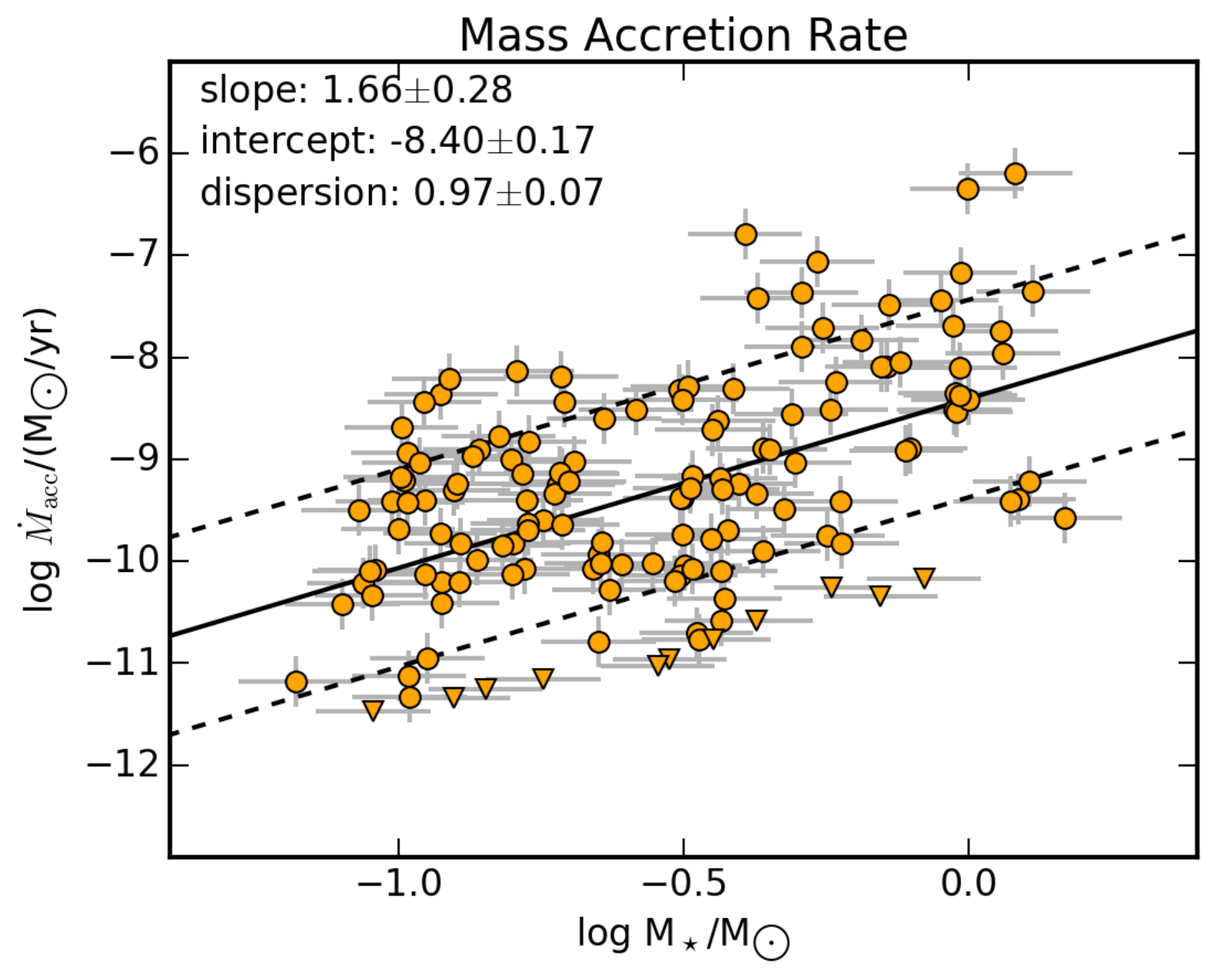}\\
	\includegraphics[width=0.47\linewidth]{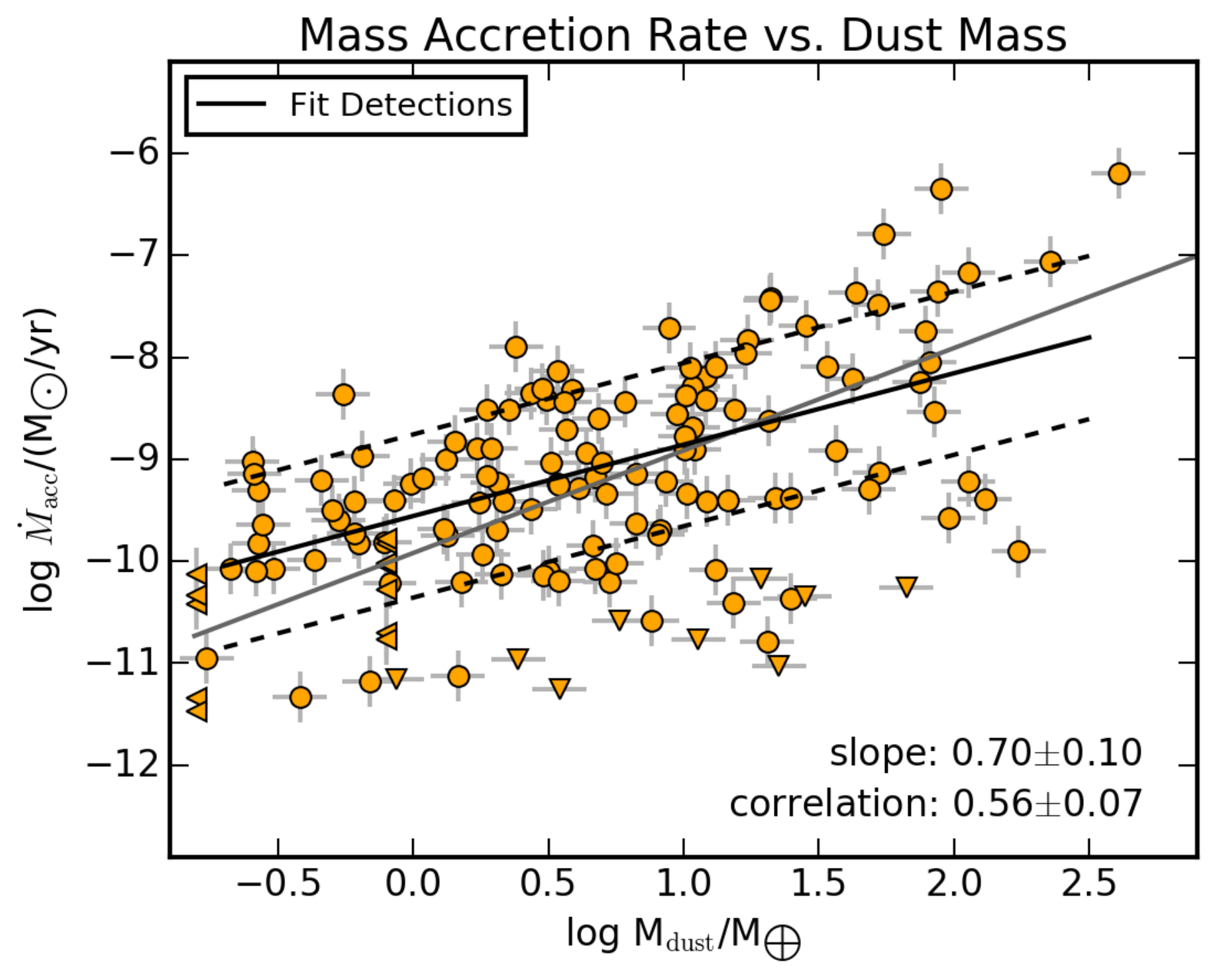}
	\includegraphics[width=0.47\linewidth]{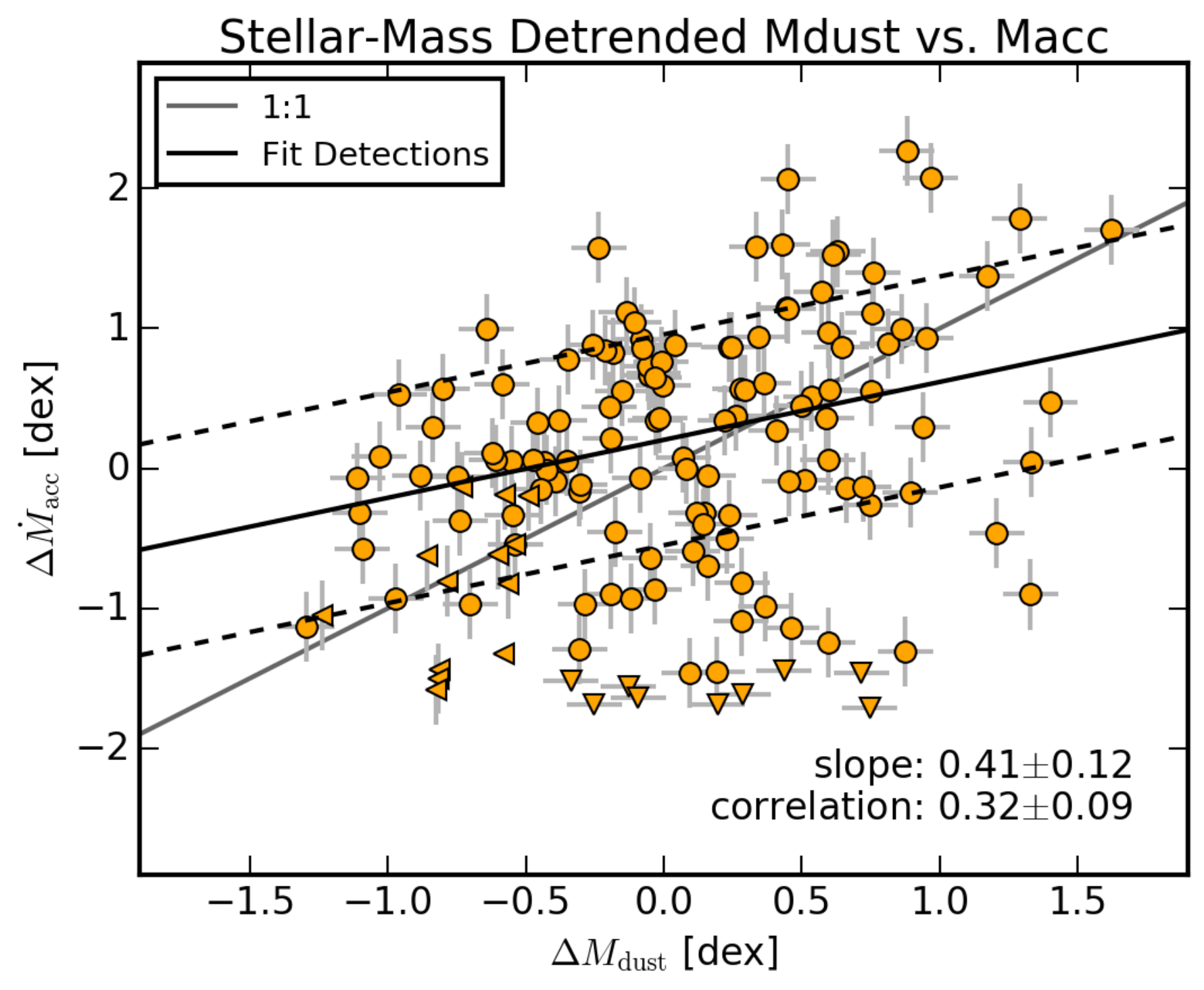}\\
      \caption{Synthetic observations for the disk model with a long viscous time scale, \visc, described in \S \ref{s:vis:3}. Same panels as Figure \ref{f:vis:1}. 
    }
    \label{f:vis:3}
\end{figure*}

This model provides a good fit to observed scatter in \Mdust and \Macc (Fig. \ref{f:vis:2}). In addition, the uncorrelated scatter weakens the observed relation between dust mass and mass accretion rate, and
these parameters provide a good fit to the observed correlation between \Mdust and \Macc ($r=0.6$) and the stellar-mass-detrended \DD--\DA correlation ($r=0.3$). We varied \facc and \fgtd independently, and found that both parameters need to be non-zero to explain the observed scatter in \Mdust and \Macc.
Although the intrinsic relation between \Mdust and \Macc in this model is linear, after applying upper limits the \Mdust--\Macc relation recovered with \linmix is shallower than linear, in agreement with the analysis of the observed values.

This model implies that (spatially unresolved) measurements of the dust mass \textit{and} (instantaneous) measurements of mass accretion rates for individual objects may not be good tracers of the disk gas. In fact, the stellar mass is a more accurate predictor of disk mass ($\sigma=0.4$ dex) compared to the measured dust mass ($\sigma=0.7$ dex).

\subsection{\visc model}\label{s:vis:3}
\cite{Lodato:prep} have recently shown that, in the framework of self-similar solutions for viscous disks, one can reproduce the shallower than linear \Mdust -- \Macc relation and the large scatter around it if most disks have not substantially evolved. 
Inspired by this work, we explore long viscous timescales in the context of the constant $\alpha$ disk model,
and determine for which input parameters and spread we can reproduce the slope and scatter in the \Mdust -- \Macc and the stellar-mass-detrended \DD--\DA  relation.

We find that the initial viscous timescale, $t_{{\rm vis},0}\propto R_{{\rm out},0}^2/\nu$, can be increased from the $\sim 0.1$ Myr in models \visa and \visb to $\sim 1$ Myr by decreasing the viscosity to $\alpha=0.001$ (or increasing the initial disk radius). Because the longer viscous time scale also reduces the mass accretion rate for a given disk mass, we increase the gas-to-dust ratio in model \visc by a factor 3 to reproduce the observed values (we discuss the implications of these choices in Sect.~\ref{s:discussion}).
The correlation between \Mdust and \Macc weakens in combination with a large dispersion in the parameters that affect most the viscous time ($R_{{\rm out},0}$, $\alpha$). For instance, a model with a dispersion in $\alpha$ of $2$ dex and outer radius of $0.5$ dex can reproduce the observed relations and scatter around them  (see \visc in Table~\ref{t:model} and Figure~\ref{f:vis:3}).

\section{Discussion}\label{s:discussion}
The modeling carried out in the previous sections points out two possible solutions to the shallower than linear \Mdust -- \Macc relation and the large scatter between these quantities.
The first possibility (\visb) is that (spatially unresolved) measurements of the dust mass \textit{and} (instantaneous) measurements of mass accretion rates for individual objects are not good tracers of the disk gas in protoplanetary disks. In \S ~\ref{disc:var} we summarize the current status on these observables and discuss ways to reduce their uncertainties. The second possibility (\visc) is that most $\sim$2-3\,Myr-old disks have not viscously evolved substantially, hence their birth properties (and scatter) remain imprinted in the observed \Mdust -- \Macc relation. We discuss in \S ~\ref{disc:longvis} the implications of this scenario and which observables are needed to test it. Finally, the constant $\alpha$ disk model may not provide a good description of disk viscosity, and we discuss various other physical processes that could contribute to the observed scatter in the \Mdust -- \Macc relation (\S ~\ref{s:otherprocess}) as well as a completely different scenario based on (MHD) disk winds (\S \ref{s:wind}).

\subsection{Mass accretion rates and disk masses: current status and possible improvements}\label{disc:var}
The mass accretion rates used in this paper are based on single-epoch observations. Accretion rate variability on different time scales will contribute to the observed scatter, as modeled through the dispersion in parameter \facc. On time scales up to a year, rotational modulation of the accretion flow by the star introduces a variability of $\sim 0.4$ dex \citep{2012MNRAS.427.1344C,2015A&A...581A..66V}.
We show in the appendix that this short-term variability is not sufficient to explain the observed scatter (\S \ref{a:vis:scatter}), as variability would need to be of order $0.7$ dex.
Other multi-epoch studies also found that accretion rate variability is smaller than the observed scatter in mass accretion rates \citep[e.g.][]{2009ApJ...694L.153N, 2013ApJS..207....5F}

Constraining accretion variability on timescales longer than a year is challenging. FU Orionis objects undergo brightening events associated with large increases in mass accretion rate, though their duty cycle is unknown and they are primarily associated with young massive disks. EXORs undergo similar brightening events but at shorter time scales. 
Large accretion rate variations have been reported on long time scales, for example the mass accretion rate of the Herbig Ae star HD 163296 has increased by $1.0$ dex in $\sim 15$ years \citep{2013ApJ...776...44M}. If such variations on decade-long time scales are common for T Tauri stars, repeated observations of accreting sources may provide a more accurate estimate of the time-average mass accretion rate. If variations in the accretion flow take place on time scales beyond that of modern astronomy ($10^2$--$10^5$ years) this may not be feasible.

The (dust) disk mass estimates from \Alma are calculated from the $887 ~\mu$m continuum flux assuming the same dust temperature, opacity, and gas-to-dust ratio for all disks. 
If these quantities vary from disk to disk, they may contribute to the observed scatter in millimeter fluxes, hence disk masses, in the following way:
\begin{itemize}
\item \textbf{Disk Size.} 
The characteristic temperature at which the disk emits depends on the spatial distribution of dust, in particular disk size (e.g. \citealt{2017ApJ...841..116H}). Spatially resolved millimeter observations show that protoplanetary disks vary in size by an order of magnitude \citep[e.g.][$\sigma\approx 0.4$ dex]{2010ApJ...723.1241A}.
In the optically thin limit ($T\propto R_{\rm disk}^{-1/2}$) these disk size variations would amount to a dispersion in millimeter fluxes of $\approx 0.2$ dex, significantly smaller than the required dispersion in \fgtd of $0.7$ dex. 
\item \textbf{Dust Opacity.}
The dust opacity at millimeter wavelengths depends on the grain size and composition \citep[e.g.][]{2006ApJ...636.1114D}. Multi-wavelength radio observations indicate there are variations between protoplanetary disks in the spectral indices, indicative of different grain size distributions \citep[e.g.][]{2010A&A...512A..15R}. These variations in grain size distributions may correspond to variations in the dust opacity by an order of magnitude, and may contribute significantly to the scatter in the observed millimeter fluxes. A better characterization of the grain size distributions using multi-wavelength observations may therefore provide a more accurate estimate of the dust disk mass.
\item \textbf{Disk Substructure.}
High-spatial resolution spatial observations indicate substructure in some protoplanetary disks that is indicative of radial drift and particle trapping \citep[e.g.][]{2013Sci...340.1199V}. 
Particle traps may be crucial in retaining a detectable amount of dust in the outer disk \citep[e.g.][]{2012A&A...538A.114P}, and the location and strength of these traps may affect dust mass estimates based on spatially unresolved observations. However, the number of spatially resolved disks is currently not large enough to asses the relevance of particle traps on the millimeter flux.
\end{itemize}

Observations of the dust continuum at high spatial resolution and multiple wavelengths for a significant number of disks may be used to provide more accurate estimates of the (dust) disk mass. 
Using the \visb model, we predict that a  
reduction in the derived uncertainty on disk mass from $0.7$ dex to $0.3$ dex should produce a detectable correlation with $r>0.5$ if disks evolve like constant $\alpha$ disks on short timescales ($<<$1\,Myr). This corresponds to a scatter around the best-fit \Mdust-\Mstar relation of $0.4$ dex, versus $0.8$ dex currently. 
Direct estimates of the gas mass for a large number of protoplanetary disks would be certainly preferable to test the \Mdisk-\Macc relation.

\subsection{Slow viscous evolution}\label{disc:longvis}
\cite{Lodato:prep} suggested a scenario according to which most $\sim$2-3\,Myr-old disks have not yet substantially evolved, and the viscosity has a steeper radial dependence than an irradiated disk where $\alpha$ is a constant. In the framework of the constant $\alpha$ disk model \visc this requires a low viscosity (or large radii at birth) in combination with a higher gas mass to enable accretion onto the star at the observed rates, since $M_{\rm disk} \propto t_{\rm vis} \Macc$. 
An implication of this model is that dust masses of protoplanetary disks are systematically underestimated by a factor of $\sim$3-10. While possible, the absolute value of the dust opacity is largely unknown \citep[e.g.][]{2000prpl.conf..533B}, this seems unlikely as it would imply that a significant fraction of the ~2-3\,Myr-old disks in Lupus and Chamaeleon~I are gravitationally unstable (see e.g. Figure~6 in \citealt{2016ApJ...831..125P}). Although these ALMA surveys are rather shallow, none of the Lupus-Chamaeleon~I disks, even the brightest and presumably most massive ones, show the spiral structures that develop in gravitationally unstable disks \citep[e.g.][]{2016PASA...33...12R,2016ARA&A..54..271K}.

Disks disperse on timescales similar to the age of Chamaeleon~I and Lupus, as evident from the decrease in the fraction of stars with a disk and with detectable accretion as cluster age increases \citep[e.g.][]{2009AIPC.1158....3M,2010A&A...510A..72F}. In the standard viscous evolution scenario, disks accrete most of their mass until star-driven photo-evaporation takes over and quickly disperses the disk, the two-timescale disk dispersal \citep[e.g.][]{2017RSOS....470114E}. Even with X-ray- and FUV-driven photo-evaporation current models estimate that the total mass lost to photo-evaporation amounts to only  $\sim$20-30\% of the initial disk mass (see Fig.~4 in \citealt{2014prpl.conf..475A}). In the slow viscous evolution scenario, disks do not lose a significant fraction of their initial mass through accretion on million-year time scales, hence even more efficient photo-evaporation or a different mechanism would be required to disperse them.

\begin{figure*}[t]
    \includegraphics[width=0.47\linewidth]{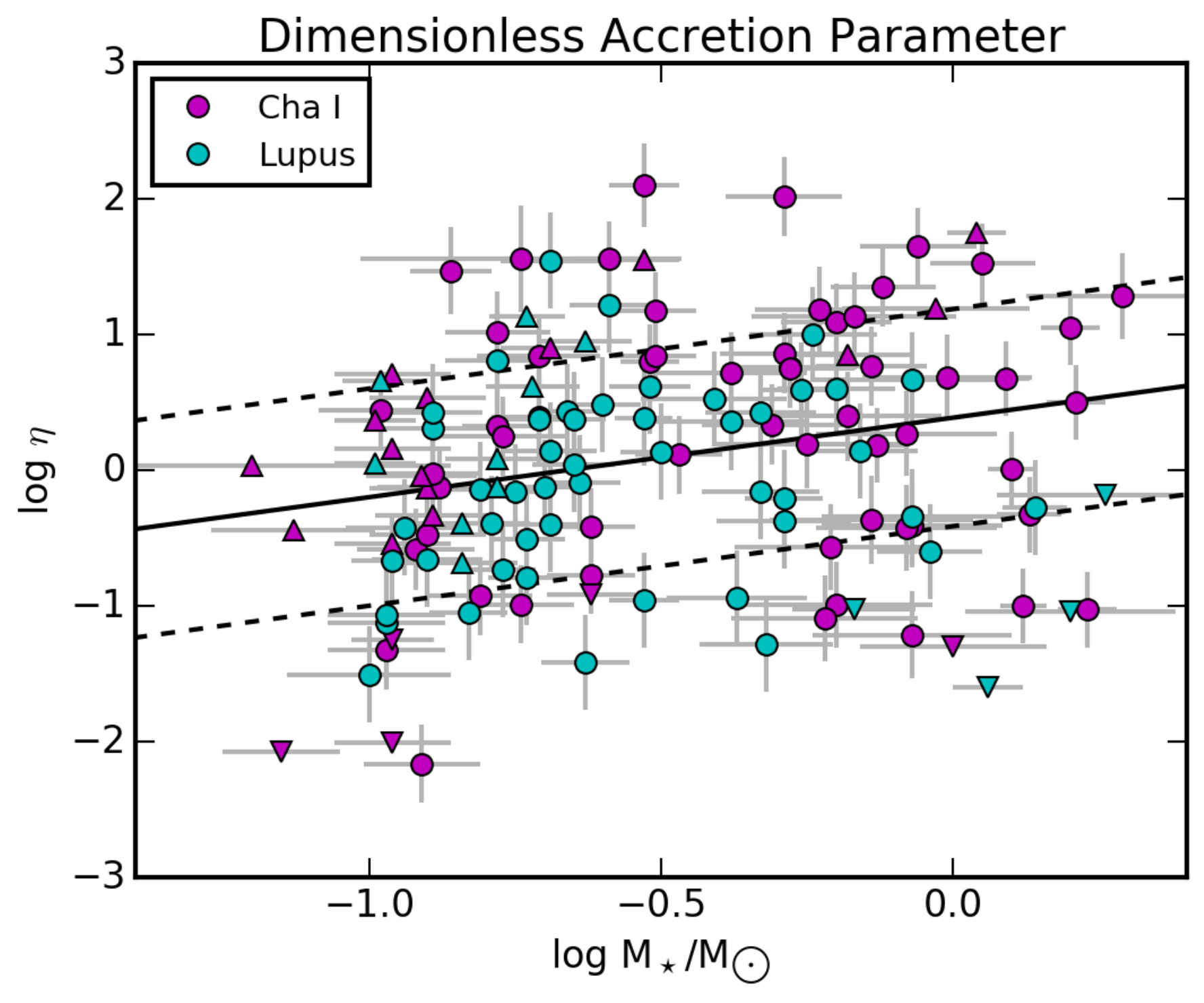}
    \includegraphics[width=0.47\linewidth]{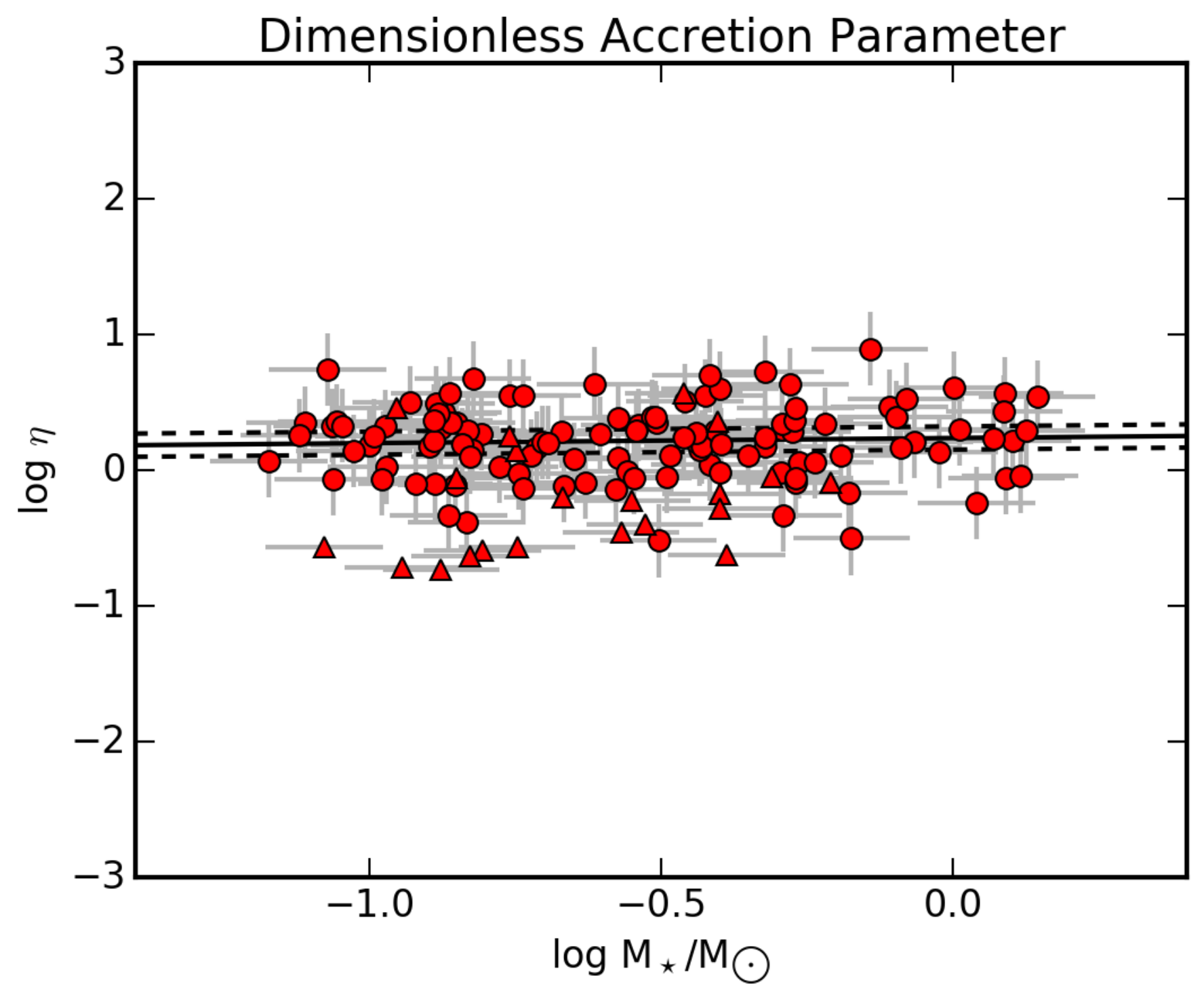}
    \caption{Dimensionless accretion parameter $\eta=\Macc ~t_{\rm disk}/(100 ~\Mdust)$ for observed disks in Lupus and Chamaeleon I (left) and for disk model \visa.
    }
    \label{f:ratio}
\end{figure*}

The evolution of mass accretion rates could provide important constraints to the slow viscous evolution scenario. Accretion rates in the Class~I stage are on average higher than those in the Class~II stage, $\sim 10^{-7}$\,M$_\odot$/yr versus $\sim 10^{-8}$\,M$_\odot$/yr \citep{2004ApJ...616..998W, 2005A&A...429..543N,2012A&A...538A..64C}. In addition, mass accretion rates of Class~II sources appear to decrease on a million-year time scale \citep{1998ApJ...495..385H,2010ApJ...710..597S,2014A&A...572A..62A,2016ARA&A..54..135H}, consistent with the standard viscous evolution scenario, but the spread is large to be certain. 
Mass accretion rate measurements for older regions, e.g. the $\sim$5-10\,Myr-old Upper Sco, would be extremely valuable as the scatter around the  \Mdust -- \Macc relation will be tighter. Evolving model \visc to $7$ Myr indicates that a correlation between \DD and \DA may become detectable at a $>3 \sigma$ level even for the long viscous time.

Finally, estimates of gas disk radii would be also important to test disk evolutionary models. In the context of viscous disk models, gas disks spread with time, hence their radii should increase. The outer radius in models \visa  and \visb  grows by a factor $\sim 20$, while the outer radius in model \visc with the longer viscous time scale grows by a factor $\sim 4$. Thus, gas disk radii as a function of class type and cluster age could directly test one of the main predictions of viscous disk models. In addition, gas disk radii, in combination with an estimate for the viscous time scale, constrain the average efficiency of angular momentum transport.

\subsection{Disk Wind}\label{s:wind}
An alternative scenario to consider is when angular momentum is not transported by turbulent viscosity but removed from the disk by an (MHD) wind \citep[e.g.][]{1982MNRAS.199..883B}. While such a scenario can be motivated on theoretical grounds and from MHD simulations \citep[e.g.][]{2016ApJ...821...80B}, quantitative predictions on how mass accretion evolves with time and for a range of conditions are missing. In particular, it is not clear how (and if) the mass accretion rate depends on disk mass and stellar mass.

Therefore, in Appendix \ref{A:wind} we construct the simplest possible wind model where the mass accretion rate depends only on the (initial) strength of the magnetic field. Such a model can fit the observed scatter in the \Mdust--\Macc plane, though we have to impose an additional stellar mass dependence in the mass accretion rate (in the $\alpha$-disk models, this dependence follows from the \Mdust--\Mstar relation). The dependence of mass accretion rate on the wind properties is likely more complex \citep[e.g.][]{2013ApJ...778L..14A}, and we stress that detailed predictions from MHD disk wind models are needed to test this scenario to the same degree as $\alpha$-disk models. The simple model is shown here only to illustrate how measurements of dust mass and mass accretion can be used to test and constrain these disk wind scenarios. 

An additional observational constraint on disk wind models is the radial extent of the gas disk. 
Winds \textit{extract} angular momentum from the system and the disk can accrete without growing in size. Viscosity, on the other hand, redistributes angular momentum within the disk and the disk grows in size when accreting. Measurements of the radial extent of the disk gas, in particular as a function of age, can provide key constraints on angular momentum transport in protoplanetary disks.

\subsection{Additional physical processes}\label{s:otherprocess}
Several physical processes, not included in the constant $\alpha$ disk model, could affect disk evolution and might contribute to the scatter in the observed \Mdisk-\Macc relation.
The impact of various physical processes on the observed dust mass and mass accretion rate were investigated by \cite{2012MNRAS.419..925J} and \cite{2017MNRAS.468.1631R}. 
Following the analysis in the latter paper, we calculate a dimensionless accretion parameter, the ratio of 
accreted mass to disk mass, 
as a measure of the accretion efficiency in the disk. Figure \ref{f:ratio} shows this ratio, here defined as
\begin{equation}
\eta=\frac{\Macc ~t_{\rm disk}}{100 ~\Mdust}
\end{equation}
disk in Lupus and Chamaeleon I (left panel) and for the constant $\alpha$-disk model \visa (right panel). The dispersion in $\eta$ is estimated by \linmix to be $0.8$ dex, 
while the \visa model predicts a much smaller dispersion of $0.1$ dex. The dispersion in $\eta$ contains similar diagnostic information as the stellar-mass detrended quantities, and the \visb and \visc model reproduce the observed dispersion in $\eta$.

The dispersion in $\eta$ can be increased by different physical processes in the following way:
\begin{itemize}
\item \textbf{Photo-evaporation.} Mass loss driven by stellar XUV photons becomes important for the disk structure when the mass accretion rate drops below the photo-evaporation rate \citep[e.g.][]{2014prpl.conf..475A}. \cite{2012MNRAS.419..925J} show that the $\eta$ increases only for a brief period at late times when the accretion rate is low, and this is unlikely to affect the dispersion in the majority of stars in our sample. External photo-evaporation increases $\eta$ by an order of magnitude \citep{2017MNRAS.468.1631R}. While there are no massive stars near the low-mass star forming regions Lupus and Chamaeleon I, external photo-evaporation might play a role in these smaller clusters under certain conditions \citep[e.g.][]{2016MNRAS.457.3593F,2017MNRAS.468L.108H}.
\item \textbf{Layered Accretion and Deadzones.} In the presence of ``dead zones'', regions with low viscosity in the mid-plane of the disk, accretion may continue through well-ionized surface layers \citep{1996ApJ...457..355G}. The build-up of material at the edge of the dead zone may trigger disk instabilities that lead to enhanced episodes of accretion. \cite{2012MNRAS.419..925J} show that layered accretion leads to a variation of orders of magnitude in $\eta$, but disks spend most of their time not accreting with only short outbursts of high accretion, which is unlikely to reproduce the observed distribution. \cite{2017MNRAS.468.1631R} suggest that a dead zone may lead to an $\eta$ below unity. Another complication with layered accretion is that it is unlikely to reproduce the stellar-mass dependence of mass accretion rates \citep{2006ApJ...648..484H}. We suggest that smaller variations in disk viscosity between surface layer and dead zone may produce a range in $\eta$ more consistent with what is observed, though constructing such a model is outside the scope of this paper.
These variations would have to be significantly smaller than those typically assumed for a dead zone ($\alpha_{\rm layer}/\alpha_{\rm dead} < 10^3-10^4$) but large enough to have a significant effect on the mass accretion rate ($>10$, see below), and could perhaps be of order $\alpha_{\rm layer}/\alpha_{\rm dead} \approx 10-100$.
\item \textbf{Radial variations in $\alpha$.} MHD simulations of protoplanetary disks find that $\alpha$ can vary radially, although variations are not large outside the dead zone, see \citep[e.g.][]{2011ApJ...735..122F,2017ApJ...835..230F}. Fig.~1 in \cite{2017MNRAS.468.1631R} shows that variations of an order of magnitude in $\alpha$ have only a negligible impact on the long-term evolution of the disk.
\item \textbf{Presence of Giant Planets.} A giant planet forming in the disk may decrease the disk accretion rate by a factor 4-10 if it is sufficiently massive \citep{2006ApJ...641..526L} and decrease the accretion efficiency $\eta$ \citep{2012MNRAS.419..925J}. Because giant planets are rare around sun-like stars ($~10\%$, \citealt{2008PASP..120..531C}) and even rarer around the low-mass stars in our sample \citep{2010PASP..122..905J,2015ApJ...814..130M} we do not expect giant planet formation to contribute significantly to the dispersion in $\eta$.
\item \textbf{Grain Growth and Radial Drift}. Grains that grow much larger than the wavelength where dust mass is estimated become undetectable, lowering $\eta$. Similarly, inward radial drift of dust grains reduces the detectable amount of dust in the outer disk, also lowering $\eta$ \citep[e.g.][]{2014prpl.conf..339T}. Dust traps are crucial in preserving a detectable amount of dust grains at millimeter wavelengths \citep[e.g.][]{2012A&A...538A.114P}. The variations in $\eta$ between disks due to grain growth and radial drift have not been quantified, but they would have to be of order $\sim 0.7 dex$ to explain the observed scatter in dust mass.
\end{itemize}

\section{Conclusions}
We analyze the \Alma dust masses and \Xshooter mass accretion rates of protoplanetary disks in the $\sim$1-3\,Myr-old Chamaeleon I and Lupus star-forming regions. 
We find that:
\begin{itemize}
\item The relation between dust mass, \Mdust, and mass accretion rate, \Macc, in Chamaeleon I has a slope consistent with linear of $0.8\pm0.2$ and a correlation coefficient of $r=0.6\pm0.1$. This result mirrors the findings in Lupus reported by \cite{2016A&A...591L...3M}.
\item There is significant scatter around the \Mdust--\Macc relationship which is not predicted by viscously evolved disk models with a constant $\alpha$. 
The scatter around the best-fit \Mdust--\Mstar relation, \DD, and the scatter around the best-fit \Macc--\Mstar, \DA, are only weakly correlated ($r\approx0.3$).
\end{itemize}

We simulate observations of an ensemble of evolving protoplanetary disks with a range of initial conditions using a Monte Carlo approach. Disk models where the viscosity is described by a constant $\alpha$ and with a viscous time scale shorter than the disk life time provide a good match to the observed \Mdust--\Mstar and \Macc--\Mstar relations. However, the predicted correlation between \Mdust--\Macc and \DD-\DA are too tight ($r>0.9$) to be consistent with the Lupus-Chamaeleon I dataset. We find two possible solutions:
\begin{enumerate}
\item The scatter in observed dust mass and mass accretion rate does not reflect a dispersion in disk initial conditions (mass, disk, $\alpha$, age). 
In this scenario, the observed scatter must arise from additional physical processes: most likely grain growth and radial drift affect the observable dust mass, while variability on large time scales affects the mass accretion rates.
These processes should introduce variations in the dust-to-gas ratio between disks with a standard deviation of $0.7$ dex {\it and} time-variability in the accretion rate with a standard deviation of $0.7$ dex, much larger than the dispersion in initial disk mass ($0.3$ dex).
\item Disks do not evolve substantially at the age of Lupus and Chamaeleon I due to a low viscosity ($\alpha\sim 0.001$) or large initial disk radius ($R_{\rm disk}>100$ au). A large dispersion in these two parameters creates scatter in the observed mass accretion rates that is not correlated with the scatter in the observed (and initial) disk mass, matching the observed weak correlation between  \DD-\DA. See also \cite{Lodato:prep}.
\end{enumerate}
The large discrepancy between the observables and 
gas disk properties indicate that dust mass and mass accretion rate may be imperfect tracers of disk evolution. 
More accurate estimates of the disk mass, for example with spatially resolved multi-wavelength observations with \Alma, and of the size of gaseous disks are critical to test different evolutionary models. 

It is also possible that a different source of angular momentum transport, such as MHD disk winds, may drive accretion in protoplanetary disks. While we show that alternate mechanisms can be consistent with the observed correlation between disk mass and mass accretion rate discussed here and in \cite{2016A&A...591L...3M}, quantitative
predictions on how wind-driven mass accretion rates scale with disk and stellar properties are needed to test this scenario.

\software{
linmix \citep{2007ApJ...665.1489K},
NumPy \citep{numpy},
Matplotlib \citep{pyplot}
}

\acknowledgments
We are grateful to the anonymous referee for a constructive review that has improved the quality of the paper.
We thank Kaitlin Kratter, Kees Dullemond, Paola Pinilla, and Mario Flock for helpful comments.
This material is based upon work supported by the National Aeronautics and Space Administration under Agreement No. NNX15AD94G for the program “Earths in Other Solar Systems”. 
The results reported herein benefited from collaborations and/or information exchange within NASA’s Nexus for Exoplanet System Science (NExSS) research coordination network sponsored by NASA’s Science Mission Directorate.

\ifhasbib
	\bibliography{paper.bbl}
\else	
	\bibliography{/Users/mulders/Dropbox/papers/papers3,/Users/mulders/Dropbox/papers/books,/Users/mulders/Dropbox/papers/software}
\fi

\appendix

\begin{figure}
    \includegraphics[width=\linewidth]{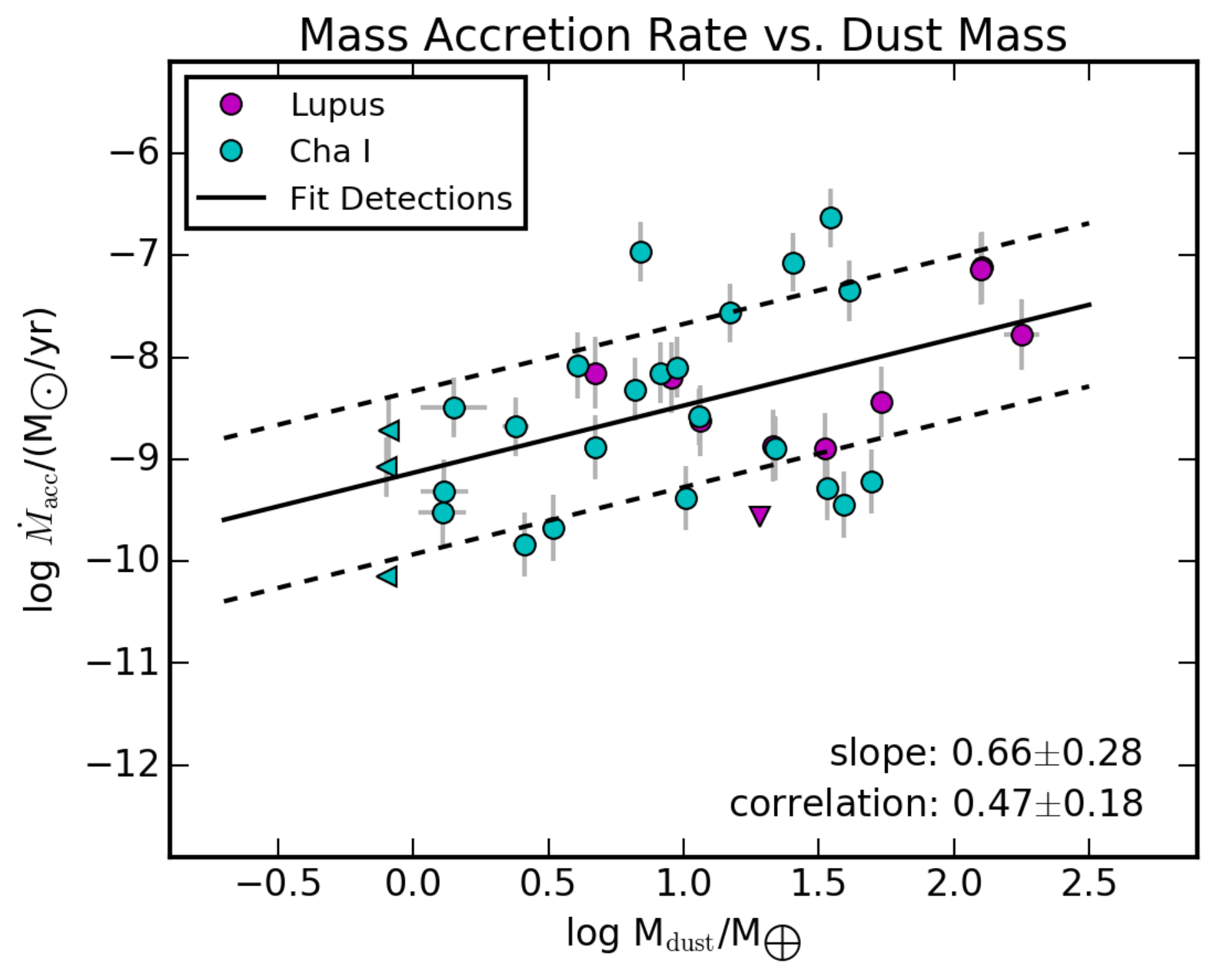}
    \caption{Dust masses versus mass accretion rates for a restricted stellar mass range, $0.5~M_\odot \leq M_\star \leq 1.0~M_\odot$. The black lines show the best-fit regression curve (solid line) and $1\sigma$ dispersion (dotted line). 
    }
    \label{f:highmass}
\end{figure}

\section{Limited stellar mass range}\label{s:limited}
The analysis of disk properties over a wide range of stellar masses may be impacted by systematic uncertainties in stellar evolutionary models or more complex stellar-mass dependencies that are not accounted for in this work. In particular, derived stellar masses suffer from larger uncertainties towards the lower mass end. On the other hand, mass accretion rates of stars more massive than a solar mass may be less reliable.  In addition, \cite{2017A&A...600A..20A} and \cite{2017arXiv170402842M} find tentative evidence for a different slope in the \Macc-\Mstar distribution at lower stellar masses. 
To assess the influence of the uncertainties described above we re-evaluate the strength of the \Macc-\Mdust correlation for a limited range in stellar masses ($0.5-1.0\Msun$) where the stellar masses and mass accretion rates are most reliable. 

Figure \ref{f:highmass} shows the mass accretion rates of the combined sample as a function of dust mass for this limited stellar mass range. We fit a moderate correlation ($r=0.5\pm0.2$) with a slope which is flatter than linear ($0.7\pm0.3$) but compatible within errors. The confidence intervals on these parameters are larger due to the lower number of stars included in the analysis. Within errors, the correlation is consistent with 
the \Macc-\Mdust 
correlation for the entire sample. There is no evidence of a tight linear correlation between \Mdust and \Macc in this restricted stellar mass range. 
Hence we conclude that there is no evidence that combining disk properties over a wide range of stellar masses reduces the strength of the observed correlations.

\begin{figure*}
   	\includegraphics[width=0.47\linewidth]{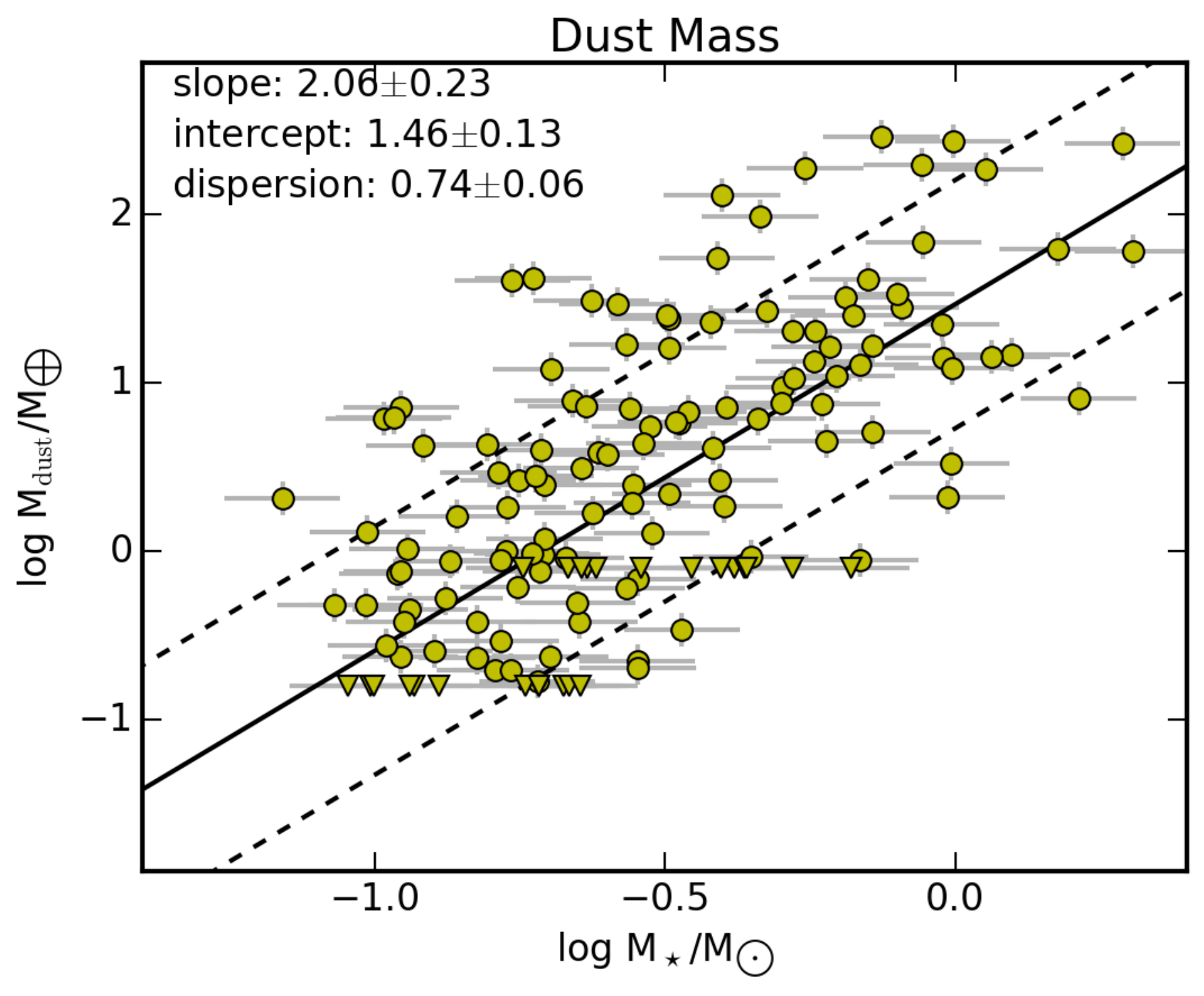}
   	\includegraphics[width=0.47\linewidth]{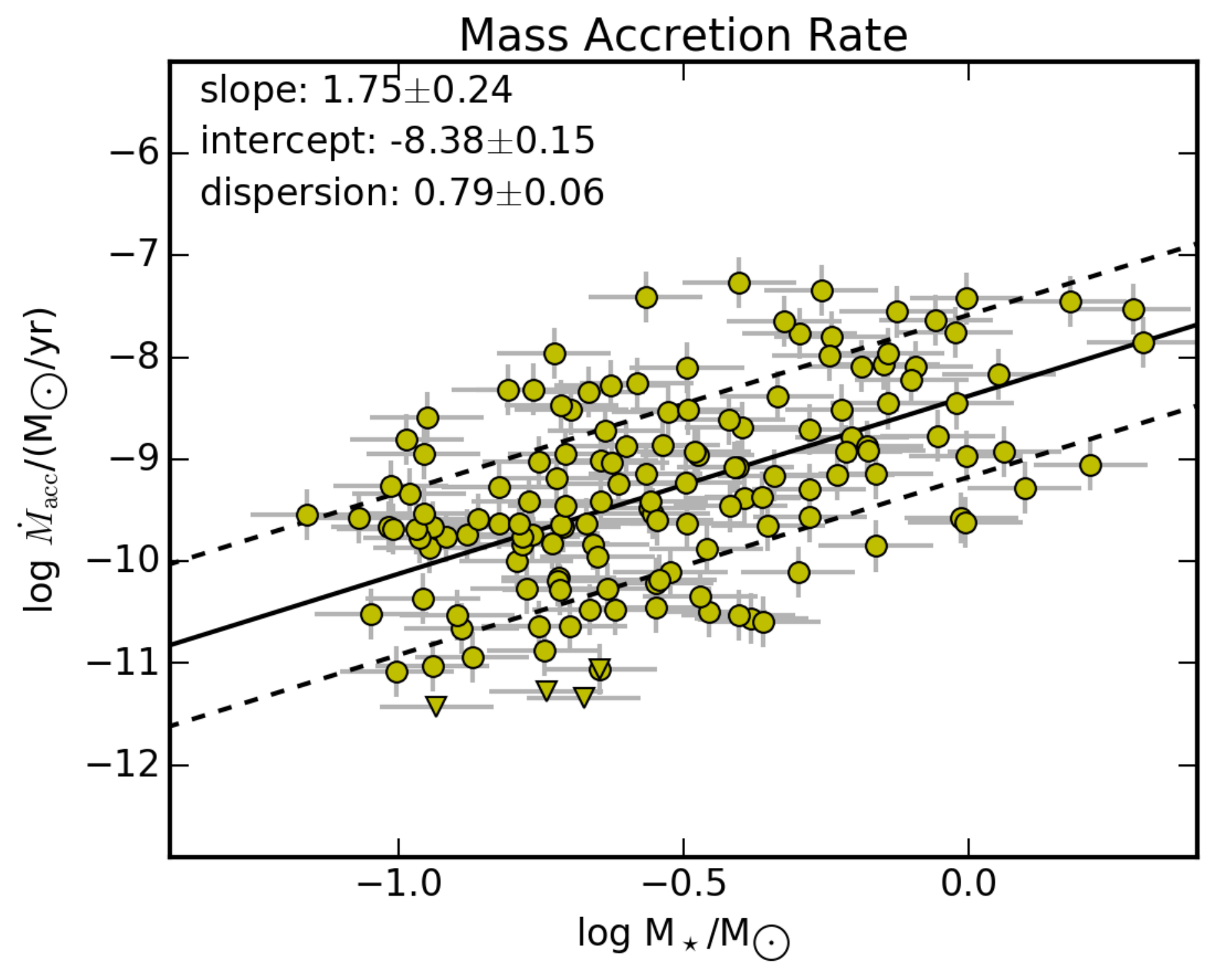}\\
	\includegraphics[width=0.47\linewidth]{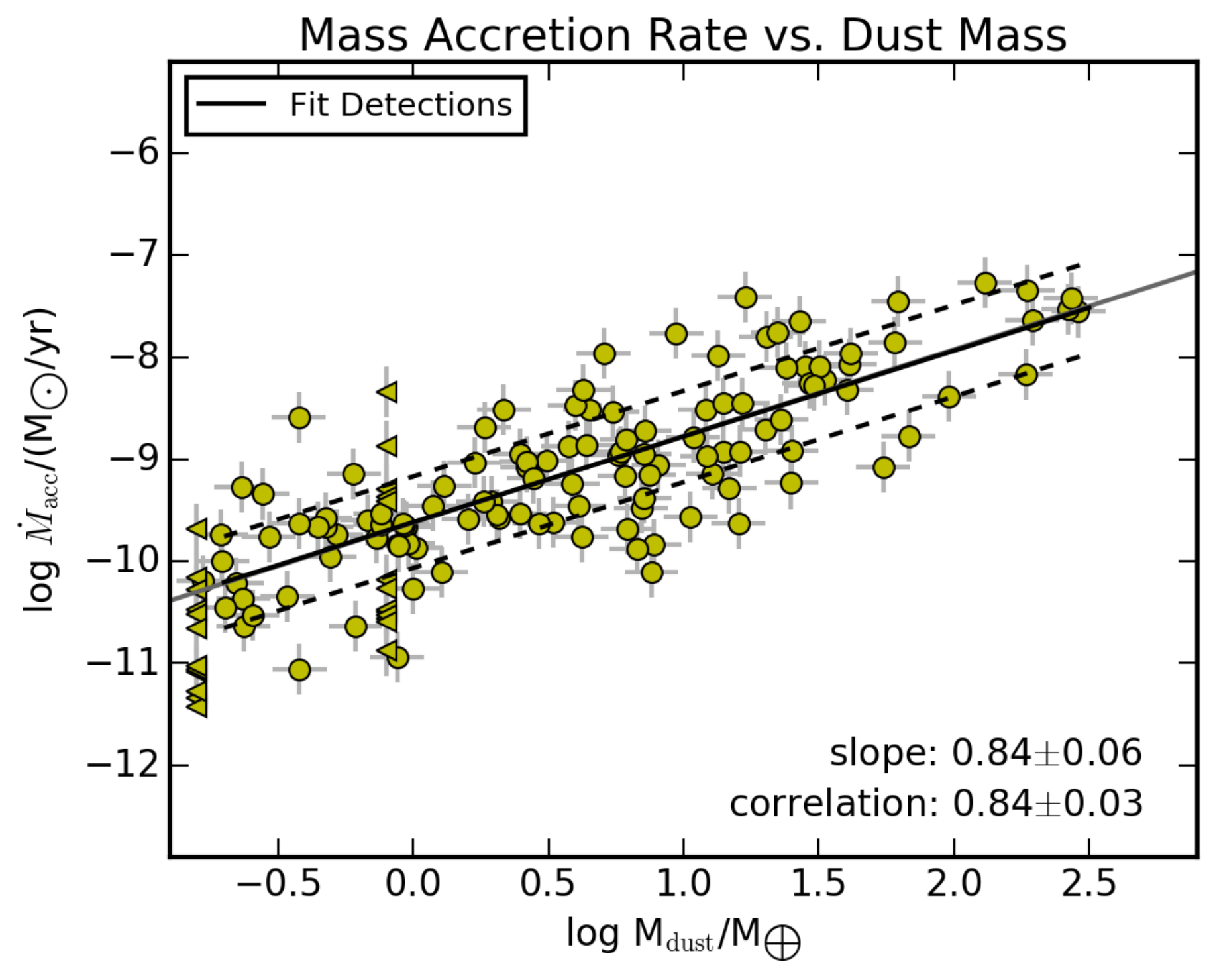}
	\includegraphics[width=0.47\linewidth]{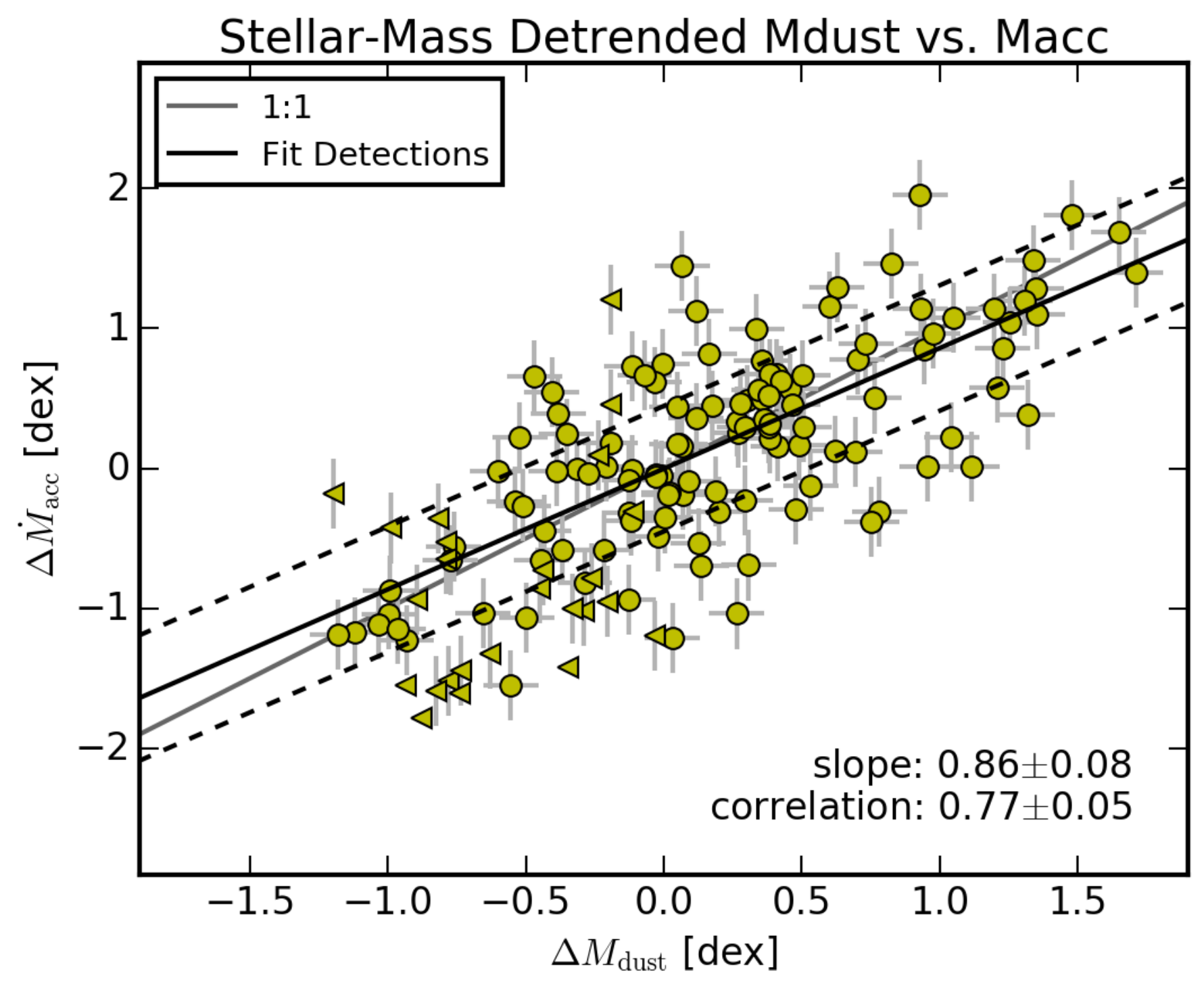}\\
      \caption{Synthetic observations for a disk model with a accretion rate variability consistent with that observed on timescales up to years. Same panels as Figure \ref{f:vis:1}. 
    }
    \label{f:vis:var}
\end{figure*}

\section{Accretion variability model}\label{a:vis:scatter}
The measured mass accretion rate is known to vary on timescales of hours to years with a typical magnitude of $\sim0.4$ dex \citep{2012MNRAS.427.1344C,2014MNRAS.440.3444C,2014A&A...570A..82V}. 
This variability would introduce additional scatter in the \Macc-\Mstar relation that could weaken the correlation between dust mass and mass accretion rate. Processes like grain growth and radial drift \citep[e.g.][]{2014ApJ...780..153B}, and disk mass-loss \citep[e.g.][]{2015ApJ...804...29G,2016ApJ...821...80B}
) change the dust-to-gas ratio of the disk, and are a potential source of scatter in the observed dust masses.

We run a set of constant $\alpha$ disk models, \visb, with a deviation between the instantaneous mass accretion rate and the disk mass accretion rate of $\facc=0.4$ dex, consistent with the observed variability. This model requires a lower dispersion in initial disk mass of $0.6$ dex to fit the observed scatter in \Macc. Because the lower dispersion in initial disk mass reduces the observable scatter in the observed dust mass as well, we introduce a dispersion in the gas-to-dust ratio of \fgtd of $0.4$ dex to fit the data. The top two panels of figure \ref{f:vis:var} show that the simulated \Mdust--\Mstar and \Macc--\Mstar are consistent with the observations. 

The added variability weakens the inferred correlation between \Mdust and \Macc ($r=0.8$), but remains more tightly correlated than observed ($r=0.6$). The difference is significant at the $2.6 \sigma$ level.
A strong correlation ($r=0.7$) remains present after detrending, indicating that the magnitude of the observed accretion rate variability on short timescales is not sufficient to erase a correlation between disk mass and mass accretion rate in the observables.

\section{Disk wind model}\label{A:wind}
An alternative approach to explaining the weak correlation between \Mdust and \Macc is assuming an evolutionary model where the mass accretion rate is not dependent on the local disk properties. This approach is motivated by recent theoretical and observational findings that non-viscous processes such as disk winds may play an important role in disk evolution as discussed in the introduction.
There are however, no quantitative predictions from MHD disk wind models of how mass accretion rate scales with stellar and disk properties. Hence, we devise a ``thought experiment'' in which we explore the observational signatures of a disk that accretes through an angular momentum transport mechanism other than a turbulent viscosity. The underlying assumption is that the (unknown) strength of the magnetic field, which varies from disk to disk, determines the mass accretion rate.

\begin{table}
	\title{Disk Wind Model Parameters}	
	\begin{tabular}{lc|c}\hline\hline
	Parameter (Unit) & Value & Dispersion (dex) \\
	\hline
	$M_{\rm disk,0}$(\Msun) 		& $0.02 M_\star^{1.9}$	& 1.0\\
	$t_{\rm disk}$(Myr)			& 2.0					& 0.3\\
	\Macc($M_{\odot /{\rm yr}}$)	& $10^{-8} M_\star^{1.9}$	& 0.8 \\		
	\hline\hline\end{tabular}
	\caption{Initial conditions for the simulated disk wind model.
	}
	\label{t:wind}
\end{table}

\begin{figure*}
   	\includegraphics[width=0.47\linewidth]{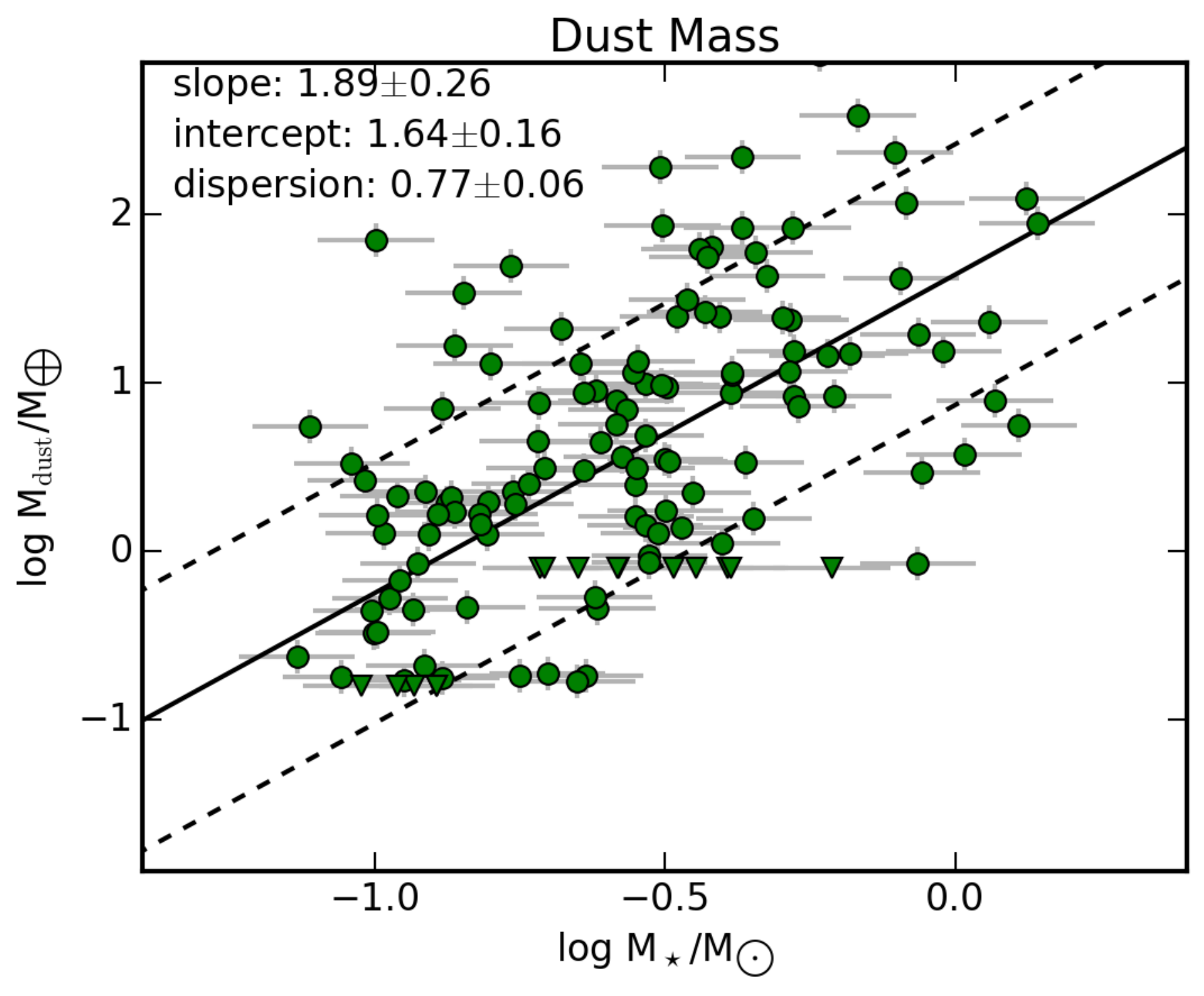}
   	\includegraphics[width=0.47\linewidth]{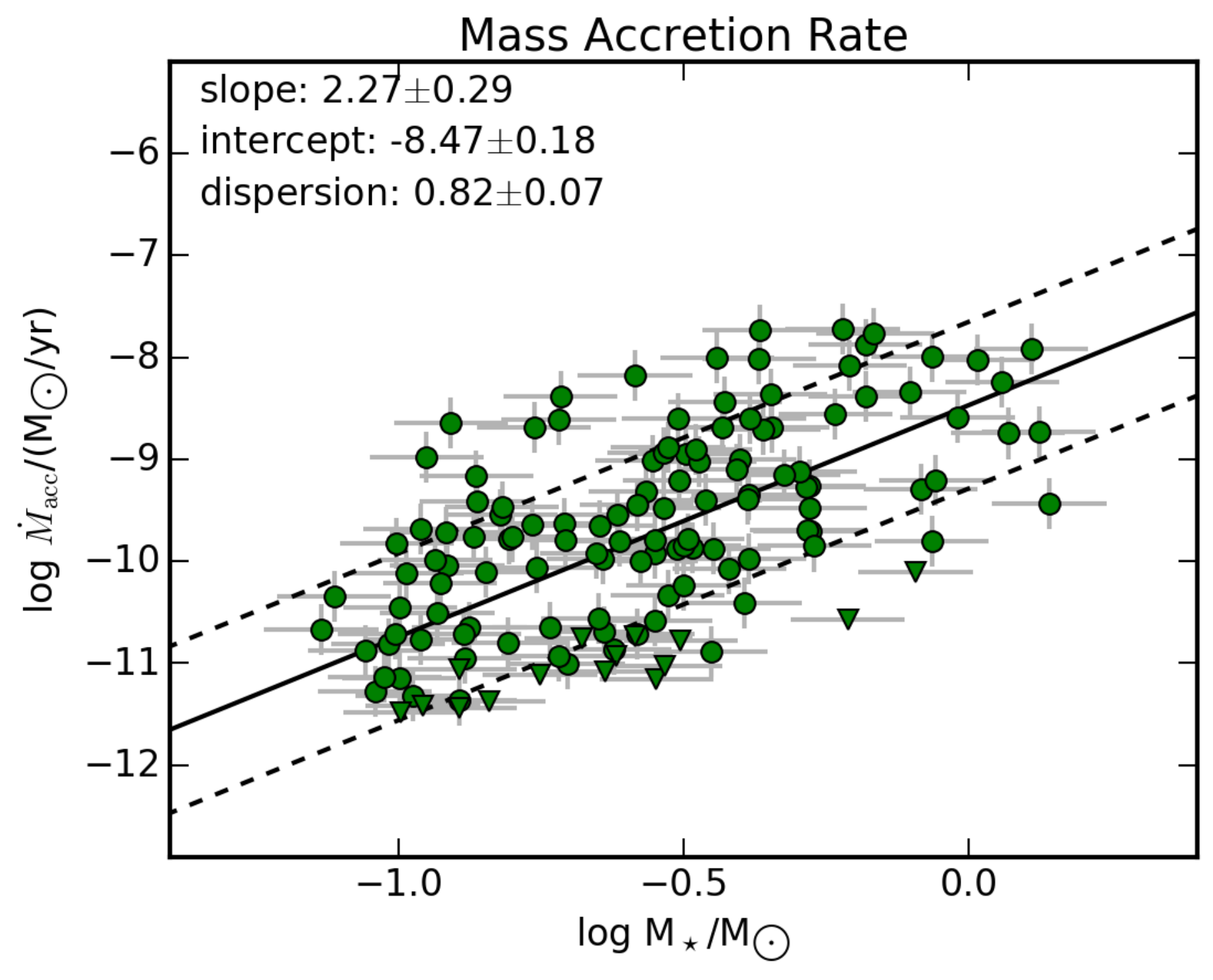}\\
	\includegraphics[width=0.47\linewidth]{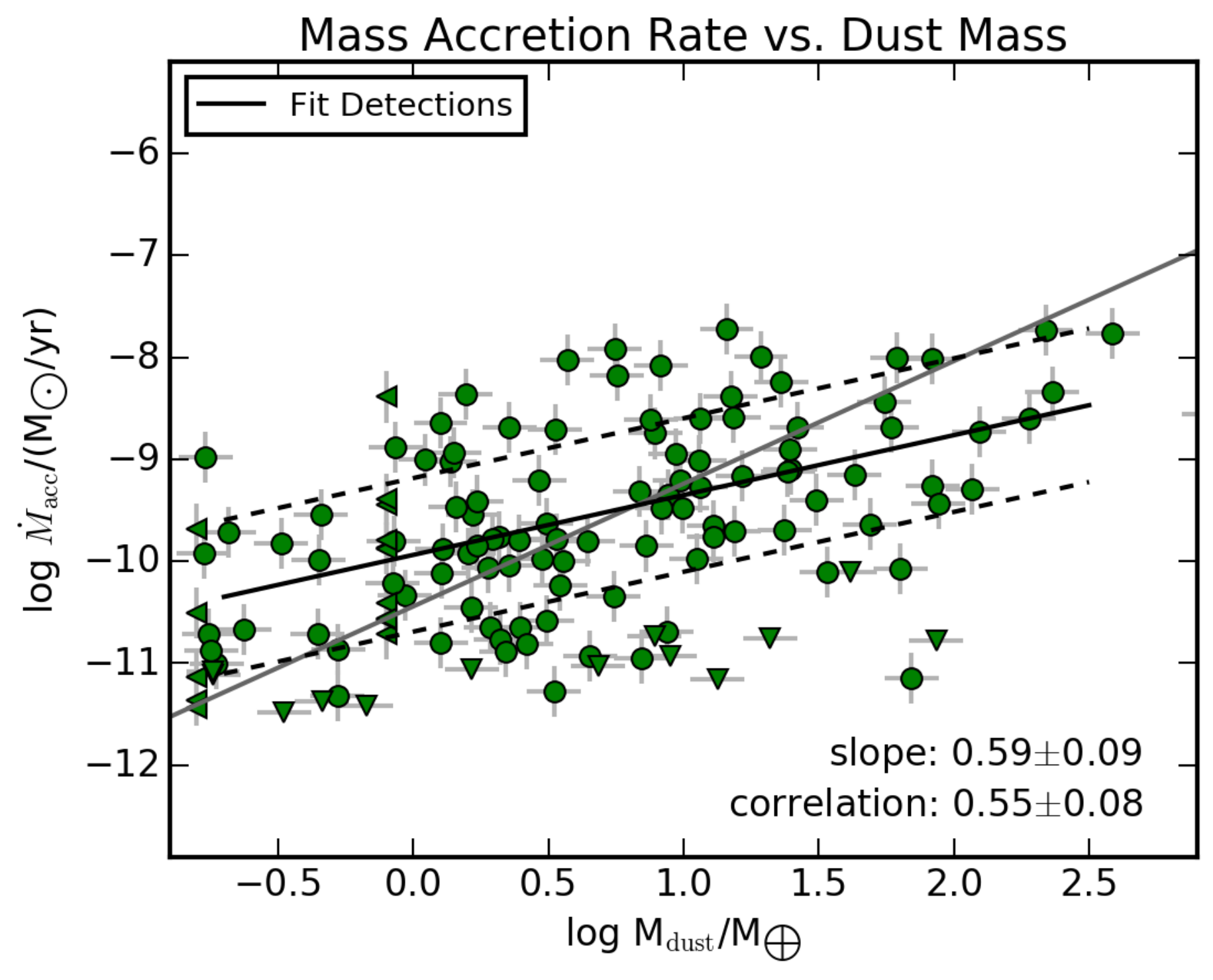}
	\includegraphics[width=0.47\linewidth]{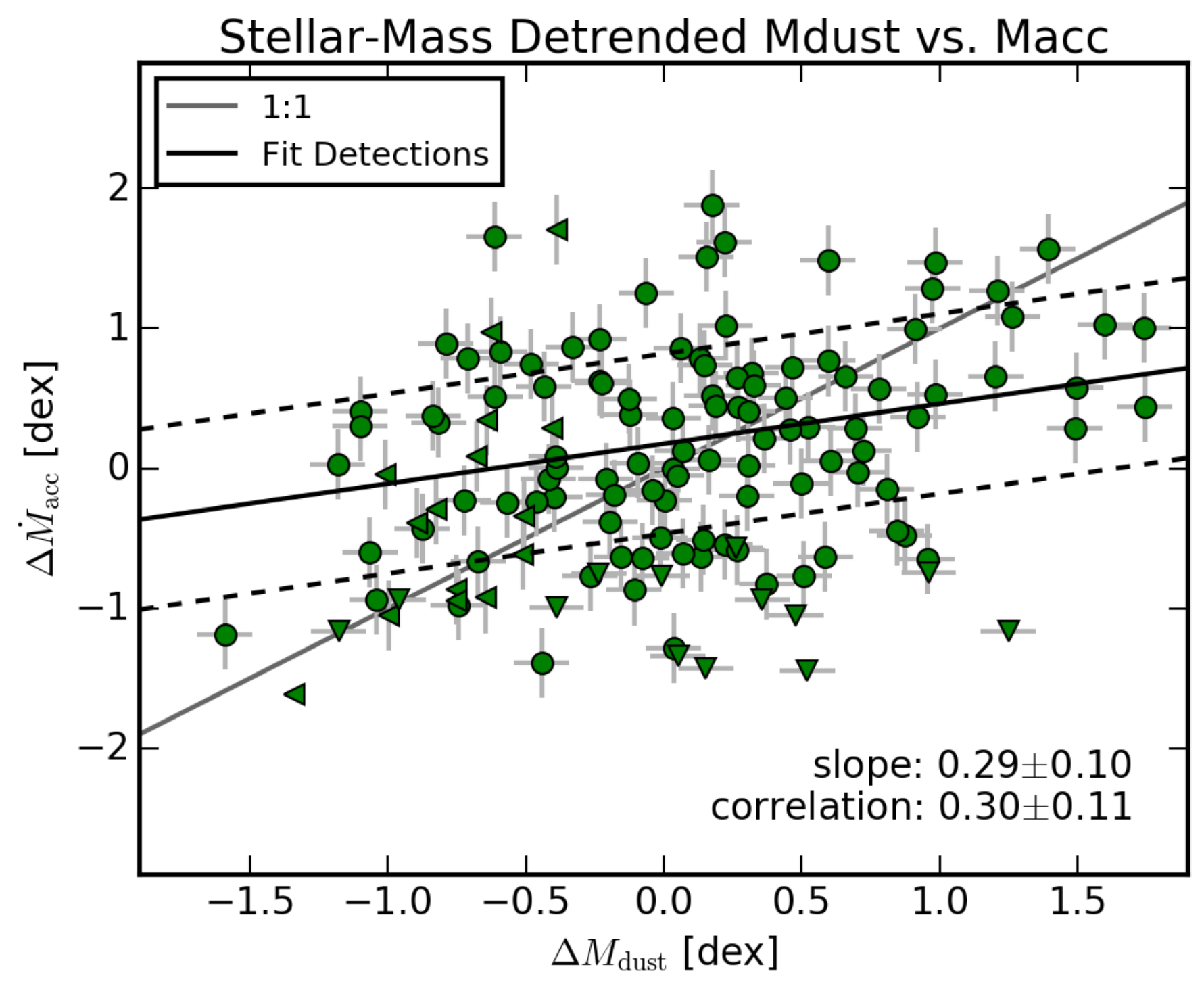}\\
    \caption{Synthetic observations for the disk wind model described in Appendix \ref{A:wind}. Same panels as Figure \ref{f:vis:1}. 
    }
    \label{f:wind}
\end{figure*}

We construct a toy model where the mass accretion rate is independent of the conditions in the disk. 
The time evolution of the disk is described by:
\begin{equation}
M_{\rm disk}=M_{\rm disk,0}- t ~\Macc.
\end{equation}
This assumption is loosely based on the constant magnetic flux model in \cite{2016ApJ...821...80B}, in which the mass accretion rate remains constant for $\sim 2$ Myr while the disk decreases in mass.
This assumption can be relaxed 
as it is not necessarily consistent with the observed decrease in mass accretion rate with age \citep[e.g.][]{2010ApJ...710..597S,2014A&A...572A..62A,2016ARA&A..54..135H}.
The disk is assumed to be dispersed if $M_{\rm disk}<0$. Disks with a larger initial mass accretion rate disperse their disks faster, leading to a lower disk mass when observed. 
The model has three main free parameters. In particular, the mass accretion rate is unconstrained by the choice of disk mass, and stellar-mass dependency has to be introduced separately in the initial disk mass and mass accretion rate to fit the observed \Mdust--\Mstar and \Macc--\Mstar relations. The free parameters in this model, $M_{\rm disk,0}$, t, \Macc, are assumed to follow a log-normal distribution. The mean disk age, $t=2$ Myr, and dispersion, $0.3$ dex, are the same as in the $\alpha$-disk model for consistency. The mean and standard deviations of the remaining two free parameters, $M_{\rm disk,0}$ and \Macc, are chosen to reproduce the observed stellar-mass dependencies and scatter in \Mdust and \Macc and listed in Table \ref{t:wind}.

We conduct synthetic observations in the same manner as for the $\alpha$-disk models, perturbing \Mstar, \Mdust, and \Macc with their respective observational uncertainties. We increase the sample size to 250 disk such that 140 disks remain at the time of observation.
The initial disk mass, $M_{\rm disk,0}$, and mass accretion rate, \Macc, are constrained by comparing the model to the observed dust mass and mass accretion rates. Note that both parameters have a stellar-mass dependency of $M_\star^{1.9}$ whereas in the $\alpha$-disk model, only the initial disk mass is assumed to be stellar-mass dependent. 
The initial distribution of disk mass accretion rates, \Macc, has a dispersion of $1.0$ dex. This yields an observable dispersion in mass accretion rates of $0.8$ dex after $2$ Myr -- consistent with the observed scatter around the \Macc--\Mstar -- and a dispersion in dust disk mass of $\sim 0.3$ dex -- significantly smaller than the observed scatter of $0.8$ dex. The additional scatter can be accounted for by introducing a dispersion in initial disk mass, $M_{\rm disk,0}$, of $0.7$ dex.

The wind model, shown in Figure \ref{f:wind}, produces a moderate correlation between dust mass and mass accretion rate of $0.6\pm0.1$, consistent with the observations. The inferred slope of the \Mdust--\Macc relation, $0.7\pm0.1$, is consistent with the slope derived from the observed data and from model \visc. The detrended quantities \DD and \DA show a weak correlation, $r=0.3\pm0.1$, again consistent with the observed correlation. No variations in the dust to gas ratio or accretion rate variability are required to fit the data.

\end{document}